\documentclass[a4paper,11pt]{article}
\usepackage{jheppub}
\usepackage{hyperref}
\usepackage{amsmath,amssymb,mathtools,dsfont}
\usepackage{float}

\usepackage{caption}
\usepackage{subcaption}
\usepackage{listings}
\usepackage{diagbox}

\usepackage{hieroglf}

\usepackage{sectsty}
\allsectionsfont{\boldmath}

\usepackage[dvipsnames]{xcolor}
\definecolor{carmine}{rgb}{0.58823529411,0,0.09411764705}
\definecolor{ForestGreen}{rgb}{0.29019607843,0.40392156862,0.25490196078}
\definecolor{dp1a}{rgb}{0.,0.333,0.486}
\definecolor{dp1b}{rgb}{0.,0.196,0.286}
\definecolor{dp1c}{rgb}{0.,0.475,0.686}
\definecolor{dp1d}{rgb}{0.,0.612,0.886}
\definecolor{clr1}{RGB}{73,129,164}
\definecolor{clr2}{RGB}{159,60,72}
\definecolor{clr3}{RGB}{17,49,71}

\definecolor{codegreen}{rgb}{0,0.6,0}
\definecolor{codegray}{rgb}{0.5,0.5,0.5}
\definecolor{codepurple}{rgb}{0.58,0,0.82}
\definecolor{backcolour}{rgb}{0.95,0.95,0.92}
\lstdefinestyle{mystyle}{
    backgroundcolor=\color{backcolour},   
    commentstyle=\color{codegreen},
    keywordstyle=\color{magenta},
    numberstyle=\tiny\color{codegray},
    stringstyle=\color{codepurple},
    basicstyle=\ttfamily\footnotesize,
    breakatwhitespace=false,         
    breaklines=true,                 
    captionpos=b,                    
    keepspaces=true,                 
    numbers=left,                    
    numbersep=5pt,                  
    showspaces=false,                
    showstringspaces=false,
    showtabs=false,                  
    tabsize=2
}
\usepackage{tikz} 
\usetikzlibrary{shapes,arrows,decorations.markings,calc}
\usepackage{tikz-cd}
\usepackage[outline]{contour}
\usepackage{pgfplots}
\pgfplotsset{compat=1.14}

\def\hexsidee{0.35}
\def\hexsideee{0.35}
\def\hexsideeee{0.17}
\def\hexsideeeee{0.6}
\def\newside{0.14}
\def\nodept{1.2pt}

\usepackage[most]{tcolorbox}
\usepackage{orcidlink}

\title{Machine learning toric duality in brane tilings}

\author[\textpmhg{\Hibw}]{Pietro Capuozzo,}
\author[\textpmhg{\HGxvii}]{Tancredi Schettini Gherardini,}
\author[\textpmhg{\Hibw}]{and Benjamin Suzzoni}

\affiliation[\textpmhg{\Hibw}]{Southampton Theory, Astrophysics and Gravity (STAG) Research Centre \\ University of Southampton, University Road, Southampton SO17 1BJ, U.K.}
\affiliation[\textpmhg{\HGxvii}]{Centre for Theoretical Physics, School of Physical and Chemical Sciences \\ Queen Mary University of London, 327 Mile End Road, London E1 4NS, U.K.}

\emailAdd{p.capuozzo@soton.ac.uk}
\emailAdd{t.schettinigherardini@qmul.ac.uk}
\emailAdd{b.suzzoni@soton.ac.uk}

\abstract{We apply a variety of machine learning methods to the study of Seiberg duality within 4d $\mathcal{N}=1$ quantum field theories arising on the worldvolumes of D3-branes probing toric Calabi-Yau 3-folds. Such theories admit an elegant description in terms of bipartite tessellations of the torus known as brane tilings or dimer models. An intricate network of infrared dualities interconnects the space of such theories and partitions it into universality classes, the prediction and classification of which is a problem that naturally lends itself to a machine learning investigation. In this paper, we address a preliminary set of such enquiries. We begin by training a fully connected neural network to identify classes of Seiberg dual theories realised on $\mathbb{Z}_m\times\mathbb{Z}_n$ orbifolds of the conifold and achieve $R^2=0.988$. Then, we evaluate various notions of robustness of our methods against perturbations of the space of theories under investigation, and discuss these results in terms of the nature of the neural network's learning. Finally, we employ a more sophisticated residual architecture to classify the toric phase space of the $Y^{6,0}$ theories, and to predict the individual gauged linear $\sigma$-model multiplicities in toric diagrams thereof. In spite of the non-trivial nature of this task, we achieve remarkably accurate results; namely, upon fixing a choice of Kasteleyn matrix representative, the regressor achieves a mean absolute error of $0.021$. We also discuss how the performance is affected by relaxing these assumptions.
}


\begin{document}
\maketitle
\flushbottom

\section{Introduction and summary}

In recent years, machine learning (ML) methods have enjoyed an increasing use in theoretical physics, establishing a deep symbiosis which has greatly benefitted both fields. Techniques from ML have been leveraged to make progress in various branches of theoretical physics, and conversely, open questions from the latter field have stimulated the development of new methods and tools in the former. The present work belongs to the ever larger research area aimed at exploring applications of artificial neural networks (NNs) to problems in string theory and supersymmetry. In particular, we will be interested in employing NNs to investigate the space of theories which arise in a certain class of holographic constructions provided by the anti-de Sitter/conformal field theory (AdS/CFT) correspondence.

The prototypical instance of the AdS/CFT correspondence consists in the conjectured duality between type IIB superstrings propagating on AdS$_5\times S^5$ with $N$ units of five-form flux threading the 5-sphere, and 4d $\mathcal{N}=4$ super Yang-Mills (SYM) with gauge algebra $A_{N-1}$ \cite{Maldacena:1997re,Witten:1998qj,Gubser:1998bc}. The duality originates from studying a stack of $N$ D3-branes embedded in $\mathbb{R}^{1,9}$. A natural generalisation is to instead consider D3-branes in $\mathbb{R}^{1,3}\times\mathcal{M}_6$, for some 6-dimensional transverse space $\mathcal{M}_6$. Since the AdS/CFT prescription requires taking a near-horizon limit of the D-brane geometry, and since all smooth manifolds are locally flat, new pairs of holographic AdS$_5$/CFT$_4$ duals can be obtained by instead taking $\mathcal{M}_6$ to be singular. In particular, the D3-branes can be placed at the tip of the cone over a positively curved Sasaki-Einstein manifold $X_5$, in which case $\mathcal{M}_6$ is a Calabi-Yau 3-fold (CY$_3$) \cite{Acharya:1998db}. In this work, we will be particularly concerned with \textit{toric} CY 3-folds. A celebrated example is that of $X_5=T^{1,1}=(SU(2)\times SU(2))/U(1)$, wherein the holographic construction results in the duality between the type IIB superstring on AdS$_5\times T^{1,1}$ and the Klebanov-Witten theory \cite{Klebanov:1998hh}, which we shall review in greater detail below. The worldvolume gauge theories of D3-branes probing such toric singularities admit a graph theoretic presentation in terms of a dimer model, or brane tiling \cite{Franco:2005rj}. Crucially to our investigation, such dimers can be recast into a Kasteleyn matrix, against which a NN can be trained to predict key properties of the underlying dimer model.

Throughout physics, dualities have been extensively employed to describe strongly coupled dynamics in terms of weakly coupled emergent degrees of freedom. Infra-red (IR) dualities are an example of such a paradigm, whereby supersymmetric gauge theories which are equivalent in the IR admit different ultra-violet (UV) completions. Such relations partition the space of supersymmetric gauge theories into universality classes of theories flowing to the same fixed point. The classical instance of such IR dualities is that due to Seiberg \cite{Seiberg:1994pq}, which is known to also act within the space of dimers; therefore, two seemingly different tilings may in fact represent the same physical theory, provided they are related to each other by a specific (and finite) set of transformations. One of the principal aims of this work is to probe to which extent NNs are able to ``learn'' this manifestation of Seiberg duality in dimer models.

As a simple arena in which to start answering this question, we begin by investigating $\mathbb{Z}_m\times\mathbb{Z}_n$ orbifolds of the conifold $\mathcal{C}$. In particular, we build and train various elementary ML architectures to predict the tuple $(m,n)$ corresponding to an input Kasteleyn matrix. A NN successfully trained in this problem would therefore be able to identify whether two $\mathcal{C}/(\mathbb{Z}_m\times\mathbb{Z}_n)$ brane tilings are connected to each other by Seiberg duality (i.e. if they are characterised by the same values of $m$ and $n$) or not. To this end, we consider a fully connected network, trained as a regressor. The current investigation should be viewed both as a proof-of-concept, showing the ability of NNs to learn to recognise Seiberg dualities, and as a concrete starting point to tackle currently unsolved problems of similar nature with ML techniques. Amongst those, we mention the extension of this work to arbitrary brane tilings, brane bricks and brane hyper-bricks, related to integrable models; additionally, the study of dimer models might connect with the recent investigations on knot theory through ML in \cite{Gukov:2020qaj}. Indeed, as a toy model, orbifolds of the conifold retain many characteristics of more generic tilings, while also being simple enough to render the generation of large training datasets relatively straightforward.

We also perform a finer-grained analysis centred on a member of the countably infinite family of $Y^{p,q}$ tilings. Namely, we attempt to train NNs to first classify the space of toric phases of such tilings, and then to predict the individual gauged linear $\sigma$-model (GLSM) multiplicities of the corresponding toric diagram. This is a rather more complicated question for a NN to answer, and indeed we will need to resort to a somewhat more sophisticated ML architecture to obtain satisfactory results, namely a residual neural network (ResNet).

In both investigations, we are able to obtain very satisfying results. More specifically, for the orbifolds of the conifold, we find the best-performing architecture to achieve R$^2=0.988$. We also test the robustness of this result against the excision of certain regions of the dataset, both in $(m,n)$ space and in depth space, which allows us to gain some more subtle insights regarding the performance of the network. For the toric phase space of $Y^{6,0}$, we find that the ResNet architecture we build is capable of achieve accuracies of essentially 100\% in a short number of epochs, both when trained as a classifier and as a regressor, provided one fixes a choice of node and path labelling in the dimer. Indeed, we find that upon performing such trivial redefinitions, the accuracy and R$^2$ results are somewhat worsened. While not particularly concerning, this decrease in performance does entice us into the use of altogether different ML techniques which would not be affected by such redefinitions, and we provide some comments in this direction to prepare future work.

It is worth noting that ML techniques have already found fruitful applications to related problems in the past few years. In \cite{Bao:2020nbi}, the authors pioneered the use of ML algorithms, specifically na\"ive Bayes classifiers and convolutional neural networks (CNNs), for the purpose of classifying Seiberg duality in the context of quiver gauge theories. (A similar work in terms of cluster algebr\ae\ appeared recently in \cite{Dechant:2022ccf}.) Here, we extend that analysis to include the action on the superpotential, which is a major facet of Seiberg duality, by replacing quivers (and their representation in terms of adjacency matrices) with brane tilings (and their associated Kasteleyn matrices). This novel perspective also allows us to more thoroughly explore the toric phase space of the families of theories we consider. A different computational approach to the study of brane tilings can be found in \cite{Seong:2023njx}, which employed the coam\oe ba projection representation, as opposed to the more direct Kasteleyn matrix description we use here. Moreover, \cite{Seong:2023njx} focused on unsupervised techniques, such as principal component analysis,  t-distributed stochastic neighbour embedding and logistic regression, whereas we will exclusively employ supervised methods. For the interested reader, we include a few references on other applications of ML techniques to similar topics in string theory: \cite{Halverson:2019tkf, Loges:2021hvn, PhysRevD.109.106006, Arias-Tamargo:2022qgb, Loges:2022mao, Cheung:2022itk, Chen:2023whk, Alawadhi:2023gxa, Betzler_2020}.

\bigskip

This paper is structured as follows. In section~\ref{sec:review}, we review some of the theoretical background which forms the foundations of this paper. In particular, we review how brane tilings, and more generally field theories admitting a bipartite representation, arise on the worldvolumes of branes probing toric CY singularities. We then comment on the rich web of dualities which intertwines the space of such theories, before delving into a more specialised discussion on the particular families of theories that we will focus on in this work. In section~\ref{sec:results0}, we describe in greater detail the problems on which we will be training our NNs, as well as discuss our precise investigation pipeline, from data generation to network architecture. Here, we also present the results of our various investigations. Finally, in section~\ref{ref:conclusions}, we summarise our main points and sketch a number of possible extensions and generalisations of this work.

\section{Review}\label{sec:review}

In this section, we summarise the theoretical background underlying this work. We start by reviewing elementary details regarding branes probing singularities and how the dimer description can capture the physics which arises on their worldvolumes. We spell out some of the main features of generic dimer models, before specialising to the $\mathbb{Z}_m\times\mathbb{Z}_n$ orbifolds of the conifold and $Y^{p,q}$ theories, which are the subject of our investigations.

\subsection{\texorpdfstring{Bipartite worldvolume theories on D3-branes probing toric CY$_3$ singularities}{Bipartite worldvolume theories on D3-branes probing toric CY3 singularities}}\label{sec:reviewbft}

Superstring and M-theory branes probing singular ambient spaces of various dimensions have proved \cite{Morrison:1998cs} to be a fertile ground for harvesting new pairs of gravity/gauge duals via the AdS/CFT correspondence \cite{Maldacena:1997re,Witten:1998qj,Gubser:1998bc}. In the most elemental of such constructions, Dirichlet branes (D-branes) are placed on the du Val singularity at the origin of a $\mathbb{C}^2/\Gamma$ orbifold, where $\Gamma\subset\operatorname{SU}(2)$ is a finite subgroup which may \cite{Douglas:1996sw} or may not \cite{Johnson:1996py} be abelian. The resolution of this two complex-dimensional singular space is an asymptotically locally Euclidean (ALE) geometry. Via the McKay correspondence \cite{McKay}, such spaces are indeed organised in terms of the ADE classification
\begin{equation*}
\left\{
\begin{tikzpicture}[thick, scale=0.35]
\coordinate (a1) at (0,0);
\coordinate (a2) at (1,0);
\coordinate (a3) at (2,0);
\coordinate (a4) at (4,0);
\coordinate (a5) at (5,0);
\coordinate (a6) at (2.5,1);
\draw (a3) -- (a1) -- (a6) -- (a5) -- (a4);
\draw[dashed] (a3) -- (a4);
\filldraw[fill=yellow] (a1) circle (5pt);
\filldraw[fill=yellow] (a2) circle (5pt);
\filldraw[fill=yellow] (a3) circle (5pt);
\filldraw[fill=yellow] (a4) circle (5pt);
\filldraw[fill=yellow] (a5) circle (5pt);
\filldraw[fill=carmine] (a6) circle (5pt);
\end{tikzpicture}~
,\quad
\vcenter{\hbox{
\begin{tikzpicture}[thick, scale=0.35]
\coordinate (a1) at (0.5,1);
\coordinate (a2) at (1,0);
\coordinate (a3) at (3,0);
\coordinate (a4) at (3.5,1);
\coordinate (a5) at (3.5,-1);
\coordinate (a6) at (0.5,-1);
\draw (a1) -- (a2);
\draw (a6) -- (a2);
\draw[dashed] (a2) -- (a3);
\draw (a4) -- (a3) -- (a5);
\filldraw[fill=yellow] (a1) circle (5pt);
\filldraw[fill=yellow] (a2) circle (5pt);
\filldraw[fill=yellow] (a3) circle (5pt);
\filldraw[fill=yellow] (a4) circle (5pt);
\filldraw[fill=yellow] (a5) circle (5pt);
\filldraw[fill=carmine] (a6) circle (5pt);
\end{tikzpicture}}}~
,\quad
\begin{tikzpicture}[thick, scale=0.35]
\coordinate (a1) at (0,0);
\coordinate (a2) at (1,0);
\coordinate (a3) at (2,0);
\coordinate (a4) at (3,0);
\coordinate (a5) at (4,0);
\coordinate (a6) at (2,1);
\coordinate (a7) at (2,2);
\draw (a1) -- (a5);
\draw (a3) -- (a7);
\filldraw[fill=yellow] (a1) circle (5pt);
\filldraw[fill=yellow] (a2) circle (5pt);
\filldraw[fill=yellow] (a3) circle (5pt);
\filldraw[fill=yellow] (a4) circle (5pt);
\filldraw[fill=yellow] (a5) circle (5pt);
\filldraw[fill=yellow] (a6) circle (5pt);
\filldraw[fill=carmine] (a7) circle (5pt);
\end{tikzpicture}~
,\quad
\begin{tikzpicture}[thick, scale=0.35]
\coordinate (a1) at (0,0);
\coordinate (a2) at (1,0);
\coordinate (a3) at (2,0);
\coordinate (a4) at (3,0);
\coordinate (a5) at (4,0);
\coordinate (a6) at (2,1);
\coordinate (a7) at (5,0);
\coordinate (a8) at (-1,0);
\draw (a8) -- (a7);
\draw (a3) -- (a6);
\filldraw[fill=yellow] (a1) circle (5pt);
\filldraw[fill=yellow] (a2) circle (5pt);
\filldraw[fill=yellow] (a3) circle (5pt);
\filldraw[fill=yellow] (a4) circle (5pt);
\filldraw[fill=yellow] (a5) circle (5pt);
\filldraw[fill=yellow] (a6) circle (5pt);
\filldraw[fill=yellow] (a7) circle (5pt);
\filldraw[fill=carmine] (a8) circle (5pt);
\end{tikzpicture}~
,\quad
\begin{tikzpicture}[thick, scale=0.35]
\coordinate (a1) at (0,0);
\coordinate (a2) at (1,0);
\coordinate (a3) at (2,0);
\coordinate (a4) at (3,0);
\coordinate (a5) at (4,0);
\coordinate (a6) at (3,1);
\coordinate (a7) at (5,0);
\coordinate (a8) at (6,0);
\coordinate (a9) at (-1,0);
\draw (a9) -- (a8);
\draw (a4) -- (a6);
\filldraw[fill=yellow] (a1) circle (5pt);
\filldraw[fill=yellow] (a2) circle (5pt);
\filldraw[fill=yellow] (a3) circle (5pt);
\filldraw[fill=yellow] (a4) circle (5pt);
\filldraw[fill=yellow] (a5) circle (5pt);
\filldraw[fill=yellow] (a6) circle (5pt);
\filldraw[fill=yellow] (a7) circle (5pt);
\filldraw[fill=yellow] (a8) circle (5pt);
\filldraw[fill=carmine] (a9) circle (5pt);
\end{tikzpicture}~
\right\}
\end{equation*}
of simply laced affine Dynkin diagrams. The worldvolume gauge theories on D-branes probing such singularities have a corresponding ADE gauge algebra, and preserve 8 real supercharges.

A further level of complexity is achieved by taking the D-branes to instead probe a three complex-dimensional singularity, which may be a Gorenstein quotient as above or a more generic algebro-geometric object. The resulting theories typically conserve only 4 supercharges \cite{Kachru:1998ys,Lawrence:1998ja}. Due to the chiral nature of four-dimensional $\mathcal{N}=1$ matter, Dynkin diagrams no longer suffice to fully capture the worldvolume physics. The appropriate recourse is to instead consider a directed graph, known as a quiver diagram. When of the finite orbit type, quivers do indeed admit an ADE classification \cite{Gabriel}, and provide a diagrammatic representation of the field content of supersymmetric gauge theories. From a graph theoretic perspective, one assigns a vector space to each node, or vertex, and a linear map to each edge, or link. The nodes are usually decorated with the dimensionality $N$ of the associated vector space. In gauge theoretic language, nodes, their labels, and the links between them correspond to vector multiplets, gauge group ranks $N$, and matter multiplets, respectively. In brane constructions, the gauge group rank corresponds to the number $N$ of coincident D-branes, while matter multiplets arise as open strings stretching across different (stacks of) D-branes. Finally, the orientation of an edge defines the target and domain nodes, with the corresponding chiral multiplet transforming in the $(\boldsymbol{N},\overline{\boldsymbol{N}})\equiv(\square,\overline{\square}) $ representation. For instance, an edge looping back to its starting node represents adjoint matter.

\begin{figure}
     \centering
     \begin{subfigure}[b]{0.4\textwidth}
         \centering
         \includegraphics[width=\textwidth]{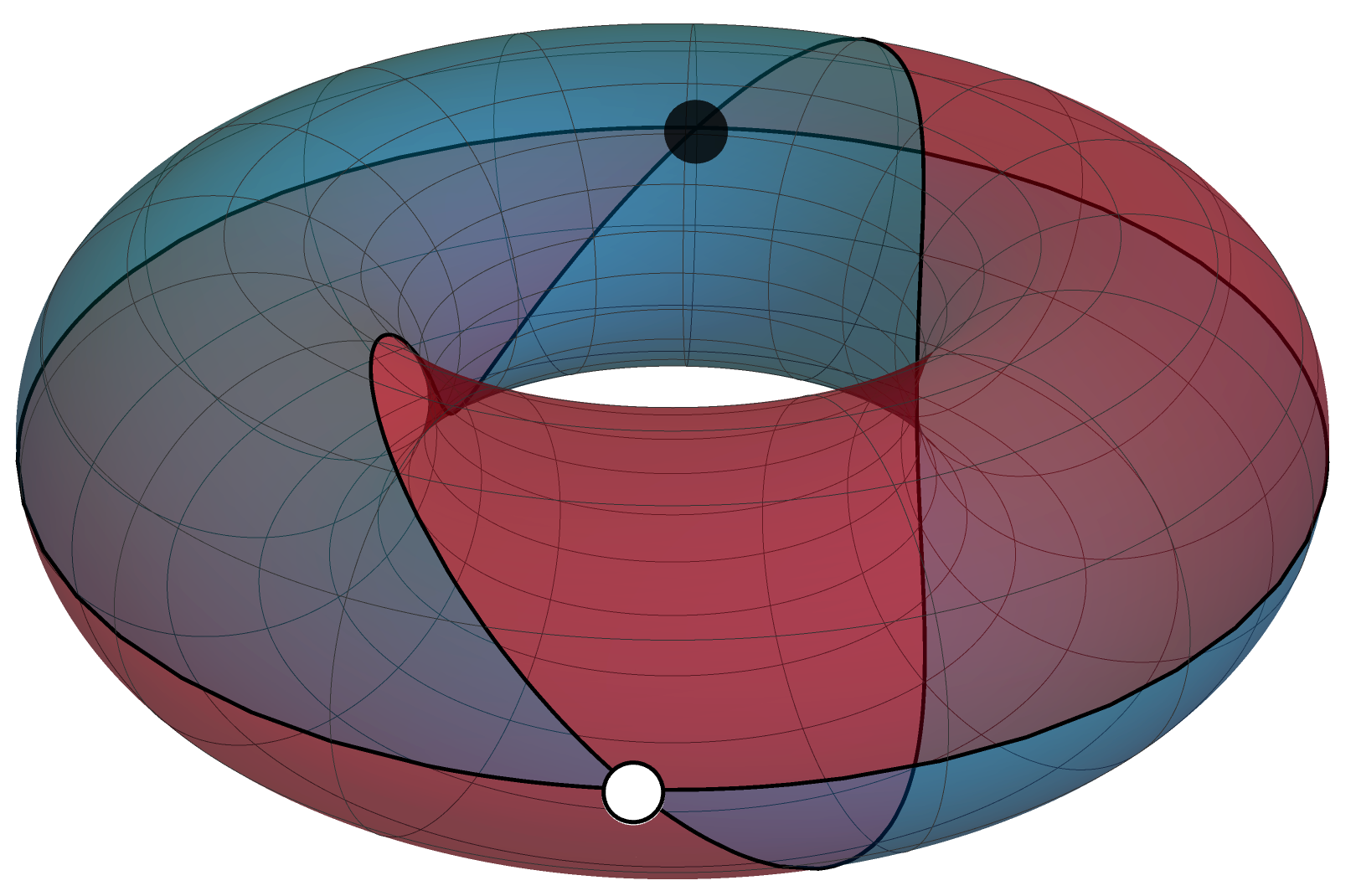}
         \caption{}
     \end{subfigure}
     \hfill
     \begin{subfigure}[b]{0.15\textwidth}
         \centering
         \begin{align*}
             \xrightarrow{\quad T^2\to\mathbb{R}^2\quad }\\ \\ \\ \\
         \end{align*}
     \end{subfigure}
     \hfill
     \begin{subfigure}[b]{0.4\textwidth}
         \centering
         \begin{tikzpicture}
            \coordinate (w1) at (0,0);
            \coordinate (w2) at (2,0);
            \coordinate (b1) at (1,0);
            \coordinate (b2) at (3,0);
            \coordinate (w3) at (1,1);
            \coordinate (w4) at (3,1);
            \coordinate (b3) at (0,1);
            \coordinate (b4) at (2,1);
            \coordinate (w5) at (0,2);
            \coordinate (w6) at (2,2);
            \coordinate (b5) at (1,2);
            \coordinate (b6) at (3,2);
            \coordinate (w7) at (4,0);
            \coordinate (w8) at (4,2);
            \coordinate (b7) at (4,1);
            \coordinate (b8) at (0,3);
            \coordinate (w9) at (1,3);
            \coordinate (b9) at (2,3);
            \coordinate (w10) at (3,3);
            \coordinate (b10) at (4,3);
            \coordinate (t1) at (0.6, 0.8);
            \coordinate (t2) at (2.6, 0.8);
            \coordinate (t3) at (1.6, 1.8);
            \coordinate (t4) at (3.6, 1.8);
            \fill[fill=carmine, opacity=0.8] (w1) -- (b1) -- (w3) -- (b3) -- cycle;
            \fill[fill=carmine, opacity=0.8] (w2) -- (b2) -- (w4) -- (b4) -- cycle;
            \fill[fill=carmine, opacity=0.8] (w3) -- (b4) -- (w6) -- (b5) -- cycle;
            \fill[fill=carmine, opacity=0.8] (w4) -- (b7) -- (w8) -- (b6) -- cycle;
            \fill[fill=carmine, opacity=0.8] (w6) -- (b6) -- (w10) -- (b9) -- cycle;
            \fill[fill=carmine, opacity=0.8] (w5) -- (b5) -- (w9) -- (b8) -- cycle;
            \fill[fill=MidnightBlue, opacity=0.8] (w2) -- (b4) -- (w3) -- (b1) -- cycle;
            \fill[fill=MidnightBlue, opacity=0.8] (w3) -- (b5) -- (w5) -- (b3) -- cycle;
            \fill[fill=MidnightBlue, opacity=0.8] (w4) -- (b6) -- (w6) -- (b4) -- cycle;
            \fill[fill=MidnightBlue, opacity=0.8] (w7) -- (b7) -- (w4) -- (b2) -- cycle;
            \fill[fill=MidnightBlue, opacity=0.8] (w6) -- (b9) -- (w9) -- (b5) -- cycle;
            \fill[fill=MidnightBlue, opacity=0.8] (w8) -- (b6) -- (w10) -- (b10) -- cycle;
            \draw (w1) -- (w7);
            \draw (b3) -- (b7);
            \draw (w5) -- (w8);
            \draw (w1) -- (b8);
            \draw (b1) -- (w9);
            \draw (w2) -- (b9);
            \draw (b2) -- (w10);
            \draw (w7) -- (b10);
            \draw (b8) -- (b10);
            \draw[dashed] (w1) -- ++(180:0.5cm);
            \draw[dashed] (b3) -- ++(180:0.5cm);
            \draw[dashed] (b8) -- ++(180:0.5cm);
            \draw[dashed] (w5) -- ++(180:0.5cm);
            \draw[dashed] (b7) -- ++(0:0.5cm);
            \draw[dashed] (w7) -- ++(0:0.5cm);
            \draw[dashed] (w8) -- ++(0:0.5cm);
            \draw[dashed] (b10) -- ++(0:0.5cm);
            \draw[dashed] (w1) -- ++(270:0.5cm);
            \draw[dashed] (w2) -- ++(270:0.5cm);
            \draw[dashed] (b1) -- ++(270:0.5cm);
            \draw[dashed] (b2) -- ++(270:0.5cm);
            \draw[dashed] (w7) -- ++(270:0.5cm);
            \draw[dashed] (b8) -- ++(90:0.5cm);
            \draw[dashed] (w9) -- ++(90:0.5cm);
            \draw[dashed] (b9) -- ++(90:0.5cm);
            \draw[dashed] (w10) -- ++(90:0.5cm);
            \draw[dashed] (b10) -- ++(90:0.5cm);
            \draw[thick, white] (t1) -- (t2) -- (t4) -- (t3) -- cycle;
            \filldraw[fill=white] (w1) circle (2pt);
            \filldraw[fill=white] (w2) circle (2pt);
            \filldraw[fill=white] (w3) circle (2pt);
            \filldraw[fill=white] (w4) circle (2pt);
            \filldraw[fill=white] (w5) circle (2pt);
            \filldraw[fill=white] (w6) circle (2pt);
            \filldraw[fill=white] (w7) circle (2pt);
            \filldraw[fill=white] (w8) circle (2pt);
            \filldraw[fill=white] (w9) circle (2pt);
            \filldraw[fill=white] (w10) circle (2pt);
            \filldraw[fill=black] (b1) circle (2pt);
            \filldraw[fill=black] (b2) circle (2pt);
            \filldraw[fill=black] (b3) circle (2pt);
            \filldraw[fill=black] (b4) circle (2pt);
            \filldraw[fill=black] (b5) circle (2pt);
            \filldraw[fill=black] (b6) circle (2pt);
            \filldraw[fill=black] (b7) circle (2pt);
            \filldraw[fill=black] (b8) circle (2pt);
            \filldraw[fill=black] (b9) circle (2pt);
            \filldraw[fill=black] (b10) circle (2pt);
          \end{tikzpicture}
         \caption{}
     \end{subfigure}
        \caption{The dimer model capturing the Klebanov-Witten theory, presented both as (a) a tiling of $T^2$ and (b) an infinite tessellation of $\mathbb{R}^2$. The fundamental domain spanned by the torus is drawn as a white outline.}
        \label{fig:conifold-dimer}
\end{figure}
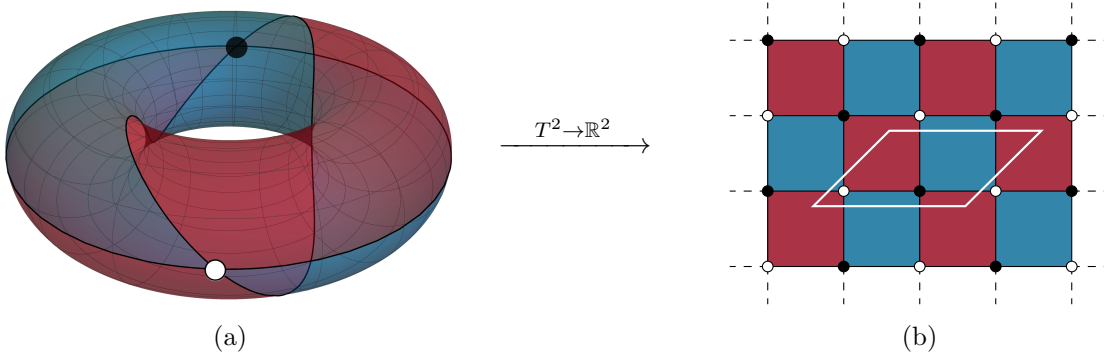

\begin{figure}
     \centering
     \begin{subfigure}[b]{0.4\textwidth}
         \centering
         \includegraphics[width=\textwidth]{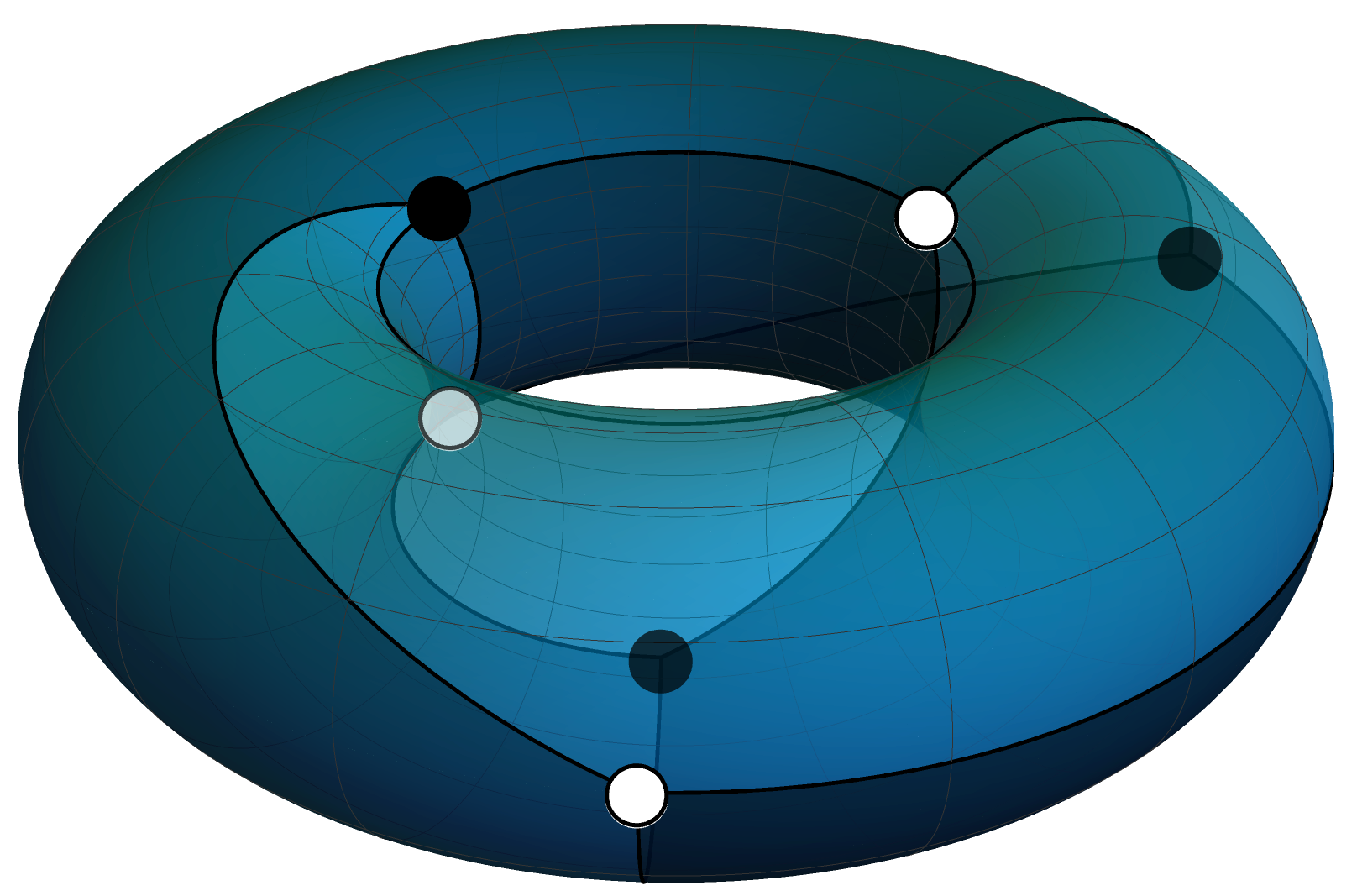}
         \caption{}
     \end{subfigure}
     \hfill
     \begin{subfigure}[b]{0.15\textwidth}
         \centering
         \begin{align*}
             \xrightarrow{\quad T^2\to\mathbb{R}^2\quad }\\ \\ \\ \\
         \end{align*}
     \end{subfigure}
     \hfill
     \begin{subfigure}[b]{0.4\textwidth}
         \centering
         \begin{tikzpicture}
            \coordinate (b1) at (0, 0);
            \coordinate (b2) at (3*\hexsideeeee, 0);
            \coordinate (b3) at (-1.5*\hexsideeeee, 0.866*\hexsideeeee);
            \coordinate (b4) at (1.5*\hexsideeeee, 0.866*\hexsideeeee);
            \coordinate (b5) at (-3*\hexsideeeee, 2*0.866*\hexsideeeee);
            \coordinate (b6) at (0, 2*0.866*\hexsideeeee);
            \coordinate (b7) at (3*\hexsideeeee, 2*0.866*\hexsideeeee);
            \coordinate (b8) at (-1.5*\hexsideeeee, 3*0.866*\hexsideeeee);
            \coordinate (b9) at (1.5*\hexsideeeee, 3*0.866*\hexsideeeee);
            \coordinate (b10) at (-3*\hexsideeeee, 4*0.866*\hexsideeeee);
            \coordinate (b11) at (0, 4*0.866*\hexsideeeee);
            \coordinate (b12) at (3*\hexsideeeee, 4*0.866*\hexsideeeee);
            \coordinate (b13) at (-1.5*\hexsideeeee, 5*0.866*\hexsideeeee);
            \coordinate (b14) at (1.5*\hexsideeeee, 5*0.866*\hexsideeeee);
            \coordinate (b15) at (-3*\hexsideeeee, 6*0.866*\hexsideeeee);
            \coordinate (b16) at (0, 6*0.866*\hexsideeeee);
            \coordinate (b17) at (3*\hexsideeeee, 6*0.866*\hexsideeeee);
            \coordinate (b18) at (-1.5*\hexsideeeee, 7*0.866*\hexsideeeee);
            \coordinate (w1) at (-1*\hexsideeeee, 0);
            \coordinate (w2) at (2*\hexsideeeee, 0);
            \coordinate (w3) at (-2.5*\hexsideeeee, 0.866*\hexsideeeee);
            \coordinate (w4) at (0.5*\hexsideeeee, 0.866*\hexsideeeee);
            \coordinate (w5) at (3.5*\hexsideeeee, 0.866*\hexsideeeee);
            \coordinate (w6) at (-1*\hexsideeeee, 2*0.866*\hexsideeeee);
            \coordinate (w7) at (2*\hexsideeeee, 2*0.866*\hexsideeeee);
            \coordinate (w8) at (-2.5*\hexsideeeee, 3*0.866*\hexsideeeee);
            \coordinate (w9) at (0.5*\hexsideeeee, 3*0.866*\hexsideeeee);
            \coordinate (w10) at (3.5*\hexsideeeee, 3*0.866*\hexsideeeee);
            \coordinate (w11) at (-1*\hexsideeeee, 4*0.866*\hexsideeeee);
            \coordinate (w12) at (2*\hexsideeeee, 4*0.866*\hexsideeeee);
            \coordinate (w13) at (-2.5*\hexsideeeee, 5*0.866*\hexsideeeee);
            \coordinate (w14) at (0.5*\hexsideeeee, 5*0.866*\hexsideeeee);
            \coordinate (w15) at (3.5*\hexsideeeee, 5*0.866*\hexsideeeee);
            \coordinate (w16) at (-1*\hexsideeeee, 6*0.866*\hexsideeeee);
            \coordinate (w17) at (2*\hexsideeeee, 6*0.866*\hexsideeeee);
            \coordinate (w18) at (-2.5*\hexsideeeee, 7*0.866*\hexsideeeee);
            \coordinate (t1) at (-1.5*\hexsideeeee-0.2, 0.866*\hexsideeeee+0.2);
            \coordinate (t2) at (1.5*\hexsideeeee-0.2, 0.866*\hexsideeeee+0.2);
            \coordinate (t3) at (-0.2, 4*0.866*\hexsideeeee+0.2);
            \coordinate (t4) at (3*\hexsideeeee-0.2, 4*0.866*\hexsideeeee+0.2);
            \draw[fill=dp1b] (w1) -- (b1) -- (w4) -- (b6) -- (w6) -- (b3) -- cycle;
            \draw[fill=dp1b] (w2) -- (b2) -- (w5) -- (b7) -- (w7) -- (b4) -- cycle;
            \draw[fill=dp1b] (w8) -- (b8) -- (w11) -- (b13) -- (w13) -- (b10) -- cycle;
            \draw[fill=dp1b] (w9) -- (b9) -- (w12) -- (b14) -- (w14) -- (b11) -- cycle;
            \draw[fill=dp1a] (w3) -- (b3) -- (w6) -- (b8) -- (w8) -- (b5) -- cycle;
            \draw[fill=dp1a] (w4) -- (b4) -- (w7) -- (b9) -- (w9) -- (b6) -- cycle;
            \draw[fill=dp1a] (w11) -- (b11) -- (w14) -- (b16) -- (w16) -- (b13) -- cycle;
            \draw[fill=dp1a] (w12) -- (b12) -- (w15) -- (b17) -- (w17) -- (b14) -- cycle;
            \draw[fill=dp1d] (w6) -- (b6) -- (w9) -- (b8) -- cycle;
            \draw[fill=dp1d] (w7) -- (b7) -- (w10) -- (b9) -- cycle;
            \draw[fill=dp1d] (w14) -- (b14) -- (w17) -- (b16) -- cycle;
            \draw[fill=dp1d] (w13) -- (b13) -- (w16) -- (b15) -- cycle;
            \draw[fill=dp1c] (b1) -- (w2) -- (b4) -- (w4) -- cycle;
            \draw[fill=dp1c] (b8) -- (w9) -- (b11) -- (w11) -- cycle;
            \draw[fill=dp1c] (b9) -- (w10) -- (b12) -- (w12) -- cycle;
            \draw[fill=dp1c] (b15) -- (w16) -- (b18) -- (w18) -- cycle;
            \draw[thick, white] (t1) -- (t2) -- (t4) -- (t3) -- cycle;
            \draw[dashed] (w1) -- ++(240:0.5cm);
            \draw[dashed] (w2) -- ++(240:0.5cm);
            \draw[dashed] (w3) -- ++(240:0.5cm);
            \draw[dashed] (b16) -- ++(60:0.5cm);
            \draw[dashed] (b17) -- ++(60:0.5cm);
            \draw[dashed] (b18) -- ++(60:0.5cm);            \draw[dashed] (w17) -- ++(120:0.5cm);
            \draw[dashed] (w18) -- ++(120:0.5cm);            \draw[dashed] (b15) -- ++(180:0.5cm);            \draw[dashed] (b10) -- ++(180:0.5cm);            \draw[dashed] (w8) -- ++(180:0.5cm);            \draw[dashed] (b5) -- ++(180:0.5cm);         \draw[dashed] (b1) -- ++(300:0.5cm);      \draw[dashed] (b2) -- ++(300:0.5cm);      \draw[dashed] (b2) -- ++(0:0.5cm);      \draw[dashed] (w5) -- ++(0:0.5cm);      \draw[dashed] (w10) -- ++(0:0.5cm);      \draw[dashed] (w15) -- ++(0:0.5cm);      \draw[dashed] (b17) -- ++(0:0.5cm);
            \filldraw[fill=black] (b1) circle (2pt);
            \filldraw[fill=black] (b2) circle (2pt);
            \filldraw[fill=black] (b3) circle (2pt);
            \filldraw[fill=black] (b4) circle (2pt);
            \filldraw[fill=black] (b5) circle (2pt);
            \filldraw[fill=black] (b6) circle (2pt);
            \filldraw[fill=black] (b7) circle (2pt);
            \filldraw[fill=black] (b8) circle (2pt);
            \filldraw[fill=black] (b9) circle (2pt);
            \filldraw[fill=black] (b10) circle (2pt);
            \filldraw[fill=black] (b11) circle (2pt);
            \filldraw[fill=black] (b12) circle (2pt);
            \filldraw[fill=black] (b13) circle (2pt);
            \filldraw[fill=black] (b14) circle (2pt);
            \filldraw[fill=black] (b15) circle (2pt);
            \filldraw[fill=black] (b16) circle (2pt);
            \filldraw[fill=black] (b17) circle (2pt);
            \filldraw[fill=black] (b18) circle (2pt);
            \filldraw[fill=white] (w1) circle (2pt);
            \filldraw[fill=white] (w2) circle (2pt);
            \filldraw[fill=white] (w3) circle (2pt);
            \filldraw[fill=white] (w4) circle (2pt);
            \filldraw[fill=white] (w5) circle (2pt);
            \filldraw[fill=white] (w6) circle (2pt);
            \filldraw[fill=white] (w7) circle (2pt);
            \filldraw[fill=white] (w8) circle (2pt);
            \filldraw[fill=white] (w9) circle (2pt);
            \filldraw[fill=white] (w10) circle (2pt);
            \filldraw[fill=white] (w11) circle (2pt);
            \filldraw[fill=white] (w12) circle (2pt);
            \filldraw[fill=white] (w13) circle (2pt);
            \filldraw[fill=white] (w14) circle (2pt);
            \filldraw[fill=white] (w15) circle (2pt);
            \filldraw[fill=white] (w16) circle (2pt);
            \filldraw[fill=white] (w17) circle (2pt);
            \filldraw[fill=white] (w18) circle (2pt);
         \end{tikzpicture}
         \caption{}
     \end{subfigure}
    \caption{The brane tiling corresponding to the $\mathbf{dP}_{\mathbf{1}}$ theory drawn on (a) the 2-torus and (b) the infinite plane.}
    \label{fig:dp1-dimer}
\end{figure}
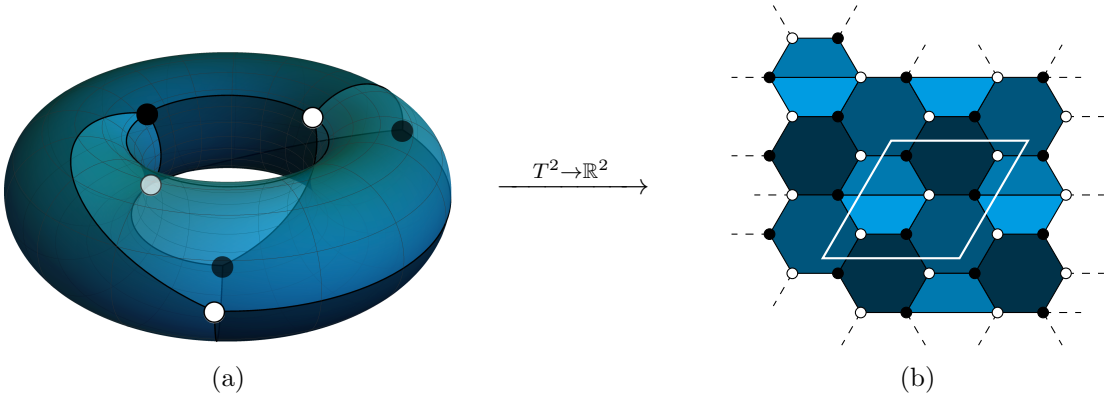

A simple and celebrated example, which will reappear frequently in the following discussion, is the quiver of the Klebanov-Witten worldvolume gauge theory on a stack of $N$ D3-branes placed at the apex of the conifold $\mathcal{C}$ \cite{Klebanov:1998hh}. The conifold corresponds to the isolated singularity $\{xy-zw=0\}\subset\mathbb{C}^4$ and describes a conical CY$_3$ over a $T^{1,1}$ base, which is a $\operatorname{U}(1)$-bundle over $S^2\times S^2$. The Klebanov-Witten quiver graph
\begin{equation}\label{eq:conifold-quiver}
    \vcenter{\hbox{
    \begin{tikzpicture}
        \coordinate (a) at (0,0);
        \coordinate (b) at (2,0);
        \begin{scope}[decoration={
            markings,
            mark=at position 0.53 with {\arrow{>>}}}
            ]
            \draw[postaction={decorate}] (a.north east) to[out=60,in=120] (b.north west);
            \draw[postaction={decorate}] (b.north east) to[out=-120,in=-60] (a.north west);
        \end{scope}
        \filldraw[fill=carmine] (a) circle (4pt);
        \filldraw[fill=MidnightBlue] (b) circle (4pt);
        \node at (-0.5,0) {$N$};
        \node at (2.5,0) {$N$};
        \node at (-2,0) {$\mathcal{C}:$};
    \end{tikzpicture}}}
\end{equation}
neatly encodes the $\operatorname{SU}(N)^2$ symmetry of the theory and the two pairs of chiral multiplets.\footnote{Throughout this work, we choose to colour each quiver node in such a way as to facilitate the identification of its corresponding face within the brane tiling of the same theory. For instance, the nodes in equations~\ref{eq:conifold-quiver} and~\ref{eq:dp1-quiver} are to be identified with the corresponding faces in figures~\ref{fig:conifold-dimer} and~\ref{fig:dp1-dimer}, respectively. However, within the quiver itself, the colours do not carry any physical information. (In the representation of a dimer as an infinite tessellation of $\mathbb{R}^2$, the colours also assist in identifying which faces are distinct from each other and which are simply repetitions of the same face.)}

Crucially, in four dimensions, $\mathcal{N}=1$ supersymmetry does not suffice to uniquely determine the Lagrangian of a theory solely in terms of the specification of a quiver diagram. Instead, the gauge and matter data represented by the quiver must be supplemented by a superpotential $\mathcal{W}$, which controls the interactions amongst the matter fields. For instance, if one identifies the conifold variables
\begin{align}
    &x=A_1B_1~,&
    &y=A_2B_2~,&
    &z=A_1B_2~,&
    &w=A_2B_1~,&
\end{align}
so that $\mathcal{C}$ is now realised as the holomorphic quotient of $\mathbb{C}^4$ by the $\mathbb{C}^*$ action
\begin{align}
    (A_i,B_j)\longmapsto\left(\lambda A_i,\lambda^{-1}B_j\right)
\end{align}
for $\lambda\in\mathbb{C}^*$, then the superpotential of the conifold theory takes the quartic form \cite{Klebanov:1998hh}
\begin{align}
    \mathcal{W}_\mathcal{C}=\varepsilon^{ij}\varepsilon^{k\ell}\operatorname{Tr}A_iB_kA_jB_\ell~.
\end{align}

\begin{figure}[t]
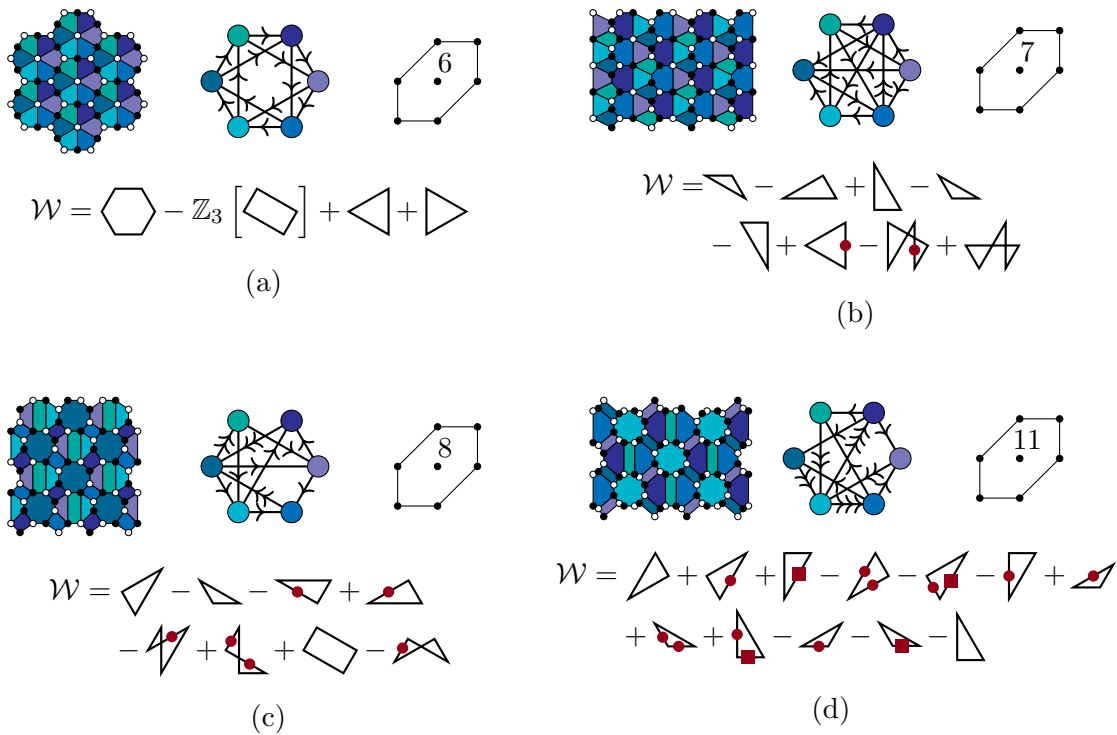

    \centering
\begin{center}
\begin{tcbraster}[raster columns=2,raster equal height]
\begin{tcolorbox}[colback=white,colframe=white,sharp corners]
        \begin{minipage}{.4\textwidth}
}}
        \end{align*}
        \begin{equation*}
            \quad\ \ \text{(b)}
        \end{equation*}
    \end{tcolorbox}
\end{tcbraster}
\begin{tcbraster}[raster columns=2,raster equal height]
\begin{tcolorbox}[colback=white,colframe=white,sharp corners]
    \begin{minipage}{.4\textwidth}
        %
}}
        \end{align*}
        \begin{equation*}
            \text{(d)}
        \end{equation*}
    \end{tcolorbox}
\end{tcbraster}
\end{center}
\caption{The four toric phases on the third del Pezzo surface, $\textbf{dP}_{\boldsymbol{3}}$. For each phase, the dimer, quiver graph, toric diagram, and superpotential are shown. This figure collects results found in \cite{Beasley:2001zp,Feng:2002zw,Franco:2005rj}. In each toric diagram, the multiplicities of the external nodes are all 1; only the multiplicity of the central node is explicitly displayed. We emphasise that, in each case, the dimer encodes all the information contained in the quiver, toric diagram, and superpotential.}
\label{fig:dp3-toric-phases}
\end{figure}

The vector space of all paths in a given quiver $Q$, equipped with the operation of concatenation, forms the complex path algebra $\mathbb{C}Q$ (for a review, see for instance \cite{Velez,broomhead2010dimermodelscalabiyaualgebras}). The superpotential $\mathcal{W}$ is then an element of the quotient $\mathbb{C}Q/[\mathbb{C}Q,\mathbb{C}Q]$ \cite{Ginzburg:2006fu}; in other words, it is a linear combination of cyclic paths. In physics parlance, it must be a function of gauge-invariant operators. Furthermore, through the F-term relations, the superpotential generates an ideal which quotients $\mathbb{C}Q$ to yield what we will refer to as the superpotential algebra. In the context of this work, the choice of a superpotential allows a quiver to be embedded into a 2-torus, $T^2$, with each term in the superpotential corresponding to a plaquette in the planar (or periodic) quiver that is produced by this construction. An important observation is that, while a theory's superpotential may indeed be graphically encoded in its quiver representation, extracting it can be a burdensome procedure -- one which usually relies on the existence of a sufficient amount of global symmetries, and occasionally on heuristic arguments. In fact, a single quiver may even admit multiple consistent superpotentials; this is the case, for instance, for the toric phase of the del Pezzo surface $\mathbf{dP}_3$ depicted in Fig.~\ref{fig:dp3-toric-phases}(b) and the $\operatorname{PdP}_{3b}$ blow-up of \cite{Feng:2004uq}. It is therefore appropriate to seek a diagrammatic representation from which superpotentials can be decoded in an unambiguous, unique and algorithmic fashion. These diagrams will turn out to be dimers, or brane tilings, and are precisely the graphs dual to periodic quivers on $T^2$.

The supersymmetric gauge theories we are interested in the present work arise in the IR description of the worldvolume physics of D3-branes probing non-compact toric Calabi-Yau 3-folds and orbifolds thereof. A toric CY$_3$ is characterised by a $\operatorname{U}(1)^3$ isometry and is realised as a metric cone over a five-dimensional Sasaki-Einstein (SE$_5$) base. The worldvolume dynamics on D3-branes probing such geometries are described by a bipartite field theory (BFT) \cite{Franco:2012mm}, a four-dimensional $\mathcal{N}=1$ quiver gauge theory which admits a definition in terms of a polygonal cell decomposition of a Riemann surface. In particular, for D3-branes probing a CY$_3$ singularity, this compact, orientable Riemann surface is the 2-torus, $T^2$, and its tessellation is referred to as a brane tiling, or dimer \cite{Franco:2005rj}. It consists of a collection of edges which, due to the 2-colourability of the graph, connect a black node with a white one, and vice versa. We restrict to balanced dimers, for which the number of black and white nodes is the same. Each face that is formed by the edges is associated to a gauge group, while an edge $i$ separating faces $a$ and $b$ represents a bifundamental chiral multiplet $\Phi_i$ transforming in $(\square_a,\overline{\square}_b)$ or $(\overline{\square}_a,\square_b)$, depending on the orientation which the edge inherits from the bipartite nature of the graph. Finally, a $k$-valent vertex contributes a term $\pm\operatorname{Tr}\Phi_{i_1}\ldots\Phi_{i_k}$ to the superpotential, the sign being determined by the colour of the node. The ordering of the chiral multiplets is deduced by going around the vertex, with the cyclicality of the trace guaranteeing that the choice of the starting edge is unphysical. In this work, we only consider leafless graphs, that is, with no univalent nodes. We can therefore take $k\geq 2$. From the dictionary spelt out above, it is clear that bivalent nodes ($k=2$) produce quadratic terms in the superpotential. Such massive fields decouple in the IR and can therefore be integrated out, which graphically corresponds to coalescing their neighbouring nodes. Nevertheless, we will usually avoid performing this operation, as to allow the algorithm to learn by itself that such operations preserve the superpotential algebra.

For instance, the brane tiling corresponding to the conifold theory can be drawn as in fig.~\ref{fig:conifold-dimer}, where we also recast the finite dimer on $T^2$ as an infinite tessellation of the plane with two-dimensional periodicity. We will frequently employ this representation of dimers in the following sections. As a richer example, one can consider the del Pezzo $\mathbf{dP}_{\mathbf{1}}$ surface, whose dimer model is shown in fig.~\ref{fig:dp1-dimer} and contains both the quiver data and the superpotential,
\begin{align}\label{eq:dp1-quiver}
\left\{\vcenter{\hbox{\begin{tikzpicture}
        \coordinate (a) at (0,0);
        \coordinate (b) at (1,0);
        \coordinate (c) at (1,1);
        \coordinate (d) at (0,1);
        \begin{scope}[decoration={
            markings,
            mark=at position 0.7 with {\arrow{>}}}
            ] 
            \draw[postaction={decorate}] (d) -- (b);
            \draw[postaction={decorate}] (d) -- (c);
            \draw[postaction={decorate}] (a) -- (c); 
        \end{scope}
        \begin{scope}[decoration={
            markings,
            mark=at position 0.6 with {\arrow{>>}}}
            ] 
            \draw[postaction={decorate}] (c) -- (b);
            \draw[postaction={decorate}] (a) -- (d);
        \end{scope}
        \begin{scope}[decoration={
            markings,
            mark=at position 0.6 with {\arrow{>>>}}}
            ] 
            \draw[postaction={decorate}] (b) -- (a);
        \end{scope}
        \filldraw[fill=dp1a] (a) circle (4pt);
        \filldraw[fill=dp1b] (b) circle (4pt);
        \filldraw[fill=dp1c] (c) circle (4pt);
        \filldraw[fill=dp1d] (d) circle (4pt);
    \end{tikzpicture}}}~,\quad
    \mathcal{W}_{\mathbf{dP}_{\mathbf{1}}}=
    \vcenter{\hbox{\begin{tikzpicture}
        \draw (b) -- (a) -- (d) node[pos=0.5, circle, fill=carmine, inner sep=2pt] {} -- cycle;
        \end{tikzpicture}}}-
        \vcenter{\hbox{\begin{tikzpicture}
        \draw (b) -- node[pos=0.5, circle, fill=carmine, inner sep=2pt] {} (a) -- (d)  -- cycle;
        \end{tikzpicture}}}-
        \vcenter{\hbox{\begin{tikzpicture}
        \draw (b) -- (a) -- (c) -- cycle node[pos=0.5, circle, fill=carmine, inner sep=2pt] {};
        \end{tikzpicture}}}+
        \vcenter{\hbox{\begin{tikzpicture}
        \draw (b) -- node[pos=0.5, circle, fill=carmine, inner sep=2pt] {} (a) -- (c) -- cycle ;
        \end{tikzpicture}}}-
        \vcenter{\hbox{\begin{tikzpicture}
        \draw (d) -- (c) -- (b) -- node[pos=0.5, rectangle, fill=carmine, inner sep=2.5pt] {}  (a) -- node[pos=0.5, circle, fill=carmine, inner sep=2pt] {} cycle ;
        \end{tikzpicture}}}+
        \vcenter{\hbox{\begin{tikzpicture}
        \draw (d) -- (c) -- node[pos=0.5, circle, fill=carmine, inner sep=2pt] {} (b) -- node[pos=0.5, rectangle, fill=carmine, inner sep=2.5pt] {}  (a) -- cycle ;
        \end{tikzpicture}}}\right\}~.
\end{align}
To avoid notational cluttering, we represent superpotentials graphically as a linear combination of closed paths in the corresponding quiver diagram. We also use $\vcenter{\hbox{\begin{tikzpicture}  \fill[carmine] (0,0) circle (4pt); \end{tikzpicture}}}$ and $\vcenter{\hbox{\begin{tikzpicture}  \fill[carmine] (0,0) rectangle (7pt,7pt); \end{tikzpicture}}}$ to denote \begin{tikzpicture}
    \begin{scope}[decoration={
            markings,
            mark=at position 0.7 with {\arrow{>>}}}
            ] 
            \draw[postaction={decorate}] (0,0) -- (0.8,0);
    \end{scope}
\end{tikzpicture} and \begin{tikzpicture}
    \begin{scope}[decoration={
            markings,
            mark=at position 0.7 with {\arrow{>>>}}}
            ] 
            \draw[postaction={decorate}] (0,0) -- (0.8,0);
    \end{scope}
\end{tikzpicture} edges. Note that each type of edge indeed appears precisely twice in the superpotential, once with either sign.

In the following work, an essential role will be played by the Kasteleyn matrix representation of dimer models. The construction of these weighted adjacency matrices proceeds as follows. One begins by assigning a weight $\pm 1$ to each edge, such that, for a face bounded by edges $\{E_1,E_2,\ldots,E_n\}$, the following holds,
\begin{equation}
    \prod_{i=1}^n\operatorname{sign}E_i=\begin{cases}
        +1 & \text{if }n=2\mod 4 \\ -1 & \text{if }n=0\mod 4~.
    \end{cases}
\end{equation}
This must be satisfied by all faces of the diagram. One then constructs two closed and oriented paths $\gamma_{w,z}$ whose winding numbers generate the homology group $H_1(T^2)$. For simplicity, we will take these paths to lie along the edges of the fundamental domain of the torus, so that their holonomies are $(0,1)$ and $(1,0)$.\footnote{While different choices of paths will in general result in different Kasteleyn matrices, the toric diagrams will be untouched up to affine transformations.} The weight of each edge is multiplied by a fugacity $w^{\pm 1}$ (resp., $z^{\pm 1}$) if it is crossed by $\gamma_w$ (resp., $\gamma_z$), where the sign in the exponent is determined by the orientation of the edge. Finally, the Kasteleyn matrix $K$ is the adjacency matrix with the weights obtained above, where each row (resp., column) corresponds to a white (resp., black) node.

\subsection{Seiberg duality, urban renewal, and toric diagrams}\label{sec:reviewseiberg}

Unlike the case of orbifold singularities \cite{Lawrence:1998ja}, there exists a large non-uniqueness of quiver gauge theories arising on D-branes probing a given toric singularity \cite{Feng:2000mi}. In particular, we will be interested in the so-called ``toric phases'', which are quiver gauge theories characterised by all gauge groups having equal rank, corresponding to the number of D-branes probing the singularity, and with each bifundamental field appearing in the superpotential in exactly one positively and one negatively signed terms. The latter is known as the toric condition, due to the fact that it follows from the toric nature of the probed singularity (which in turn implies that the F-term equations $\partial_{ij}\mathcal{W}=0$ take the form of equalities between two monomials), and is trivially satisfied given the 2-colourability of dimers. For instance, there are four distinct toric phases corresponding to the third del Pezzo $\mathbf{dP}_{\mathbf{3}}$ surface, as shown in fig. \ref{fig:dp3-toric-phases}.

In the context of the realisation of such theories via partial resolutions of Abelian orbifolds, the non-uniqueness problem can be argued by noting that unimodular transformations modify the embedding of the toric data, and therefore alter the resulting gauge theory, while of course preserving the toric variety. The physical interpretation of this phenomenon is that the extreme low-energy behaviour is indeed described by a single non-trivial RG fixed point, to which all of the distinct toric phases flow \cite{Feng:2000mi,Feng:2001xr}. The toric dualities giving rise to such universality classes are in fact a manifestation of Seiberg duality \cite{Beasley:2001zp,Feng:2001bn}.

The action of Seiberg duality at the level of dimer models has come to be known as urban renewal \cite{DBLP:journals/tcs/Propp03}, due to its local nature. In the following, we will restrict this action to quadrilateral faces only. Since toric duality maps the corresponding colour rank as $N_c\mapsto N_f-N_c$, and the flavour ranks for quadrilateral faces are given by $N_f=2N_c$, this restriction of Seiberg duality is guaranteed to act on a toric phase to produce a (generally distinct) toric phase. In particular, this restriction allows us to perform Seiberg duality an arbitrary number of times on the same quadrilateral face. This process eventually results in the creation of bivalent nodes, as discussed above, but in any case the IR physics is unchanged throughout. On the other hand, the action of Seiberg duality on non-quadrilateral faces is more subtle \cite{Franco:2005rj}; we will not take an interest in such scenarios in this paper.

At the level of the corresponding $n\times n$ Kasteleyn matrix $K$, the action of Seiberg duality is given explicitly by \cite{Franco:2005rj}
\begin{align}\label{eq:seibergduality}
    \begin{pmatrix}
        A_{(n-2)\times(n-2)} & B_{(n-2)\times 2} \\
        C_{2\times (n-2)} & \begin{matrix}
            a & b \\ c & d
        \end{matrix}
    \end{pmatrix}\longmapsto
    \begin{pmatrix}
        A_{(n-2)\times(n-2)} & B_{(n-2)\times 2} & 0_{(n-2)\times 2} \\ C_{2\times (n-2)} & 0_{2\times 2} & \begin{matrix}
            z^{d_z} & 0 \\ 0 & 1
        \end{matrix} \\ 0_{2\times(n-2)} & \begin{matrix}
            0 & -1 \\ -w^{d_w} & 0
        \end{matrix} & \begin{matrix}
            \frac{z^{d_z}}{b} & \frac{1}{d} \\ \frac{w^{d_w}z^{d_z}}{a} & \frac{w^{d_w}}{c}
        \end{matrix}
    \end{pmatrix}
\end{align}
where $d_{w,z}:=\operatorname{deg}_{w,z}abcd$, and where without loss of generality we labelled the vertices of the brane tiling such that the face which Seiberg duality acts upon lies in the lower right corner. We also assumed that the underlying brane tiling is large enough, and that the paths $\gamma_{w,z}$ are chosen such that all matrix entries are monomials -- a subtlety which we will return to later.

In the following analysis, we will identify toric phases by considering the toric diagram $\Delta_2$ associated to each dimer graph. As a necessarily rapid introduction to these objects, let us recall some rudimentary notions in toric geometry -- for more detailed reviews, see for instance \cite{fulton,Kennaway:2007tq}. We will be exclusively interested in toric varieties in three (complex) dimensions. These are algebraic varieties $V$ on which there is an action of the algebraic torus $(\mathbb{C}^*)^3\cong(U(1))^3$ -- that is, a morphism of varieties $(\mathbb{C}^*)^3\times V\longrightarrow V$ -- which has an open dense orbit on $V$. Such varieties can be combinatorially defined in terms of a fan, a collection (subject to certain closure conditions) of strongly convex rational polyhedral cones in the real vector space $N_{\mathbb{R}}=N\otimes_{\mathbb{Z}}\mathbb{R}\cong\mathbb{R}^3$, generated by elements of a lattice $N\cong\mathbb{Z}^3$. 
The fans associated to normal affine toric varieties consist of a single cone of maximal dimension (and its lower-dimensional faces). Furthermore, the Calabi-Yau condition — namely, the triviality of the canonical line bundle — translates into the cone taking the form $\Delta_2\times\{1\}\subset\mathbb{R}^3$ (up to $\operatorname{SL}(3;\mathbb{Z})$ transformations), where the toric diagram $\Delta_2\subset\mathbb{R}^2$ is a convex polytope with vertices in $\mathbb{Z}^2\subset\mathbb{R}^2$. In other words, for toric Calabi-Yau varieties, the vectors generating the toric cone are coplanar (and therefore, the variety is necessarily non-compact); such varieties can hence be defined in terms of a two-dimensional polytope $\Delta_2$.

Remarkably, it turns out \cite{Hanany:2005ve,Franco:2005rj,Franco:2006gc} that the toric diagram defined in this way is precisely the Newton polygon associated to the determinant of the Kasteleyn matrix, that is, the convex hull generated by the degrees $(d_z,d_w)$ of the various summands in\footnote{This also implies that a different choice of paths $\gamma_{w,z}$ in the dimer model amounts simply to an $\operatorname{SL}(2;\mathbb{Z})$ transformation and/or an integer shift of the associated toric diagram (i.e. an $\operatorname{SL}(3;\mathbb{Z})$ action descending onto $\Delta_2$).}
\begin{equation}
    P(z,w):=\det K=\sum_{d_z,d_w\in\Delta_2}c_{(d_z,d_w)}z^{d_z}w^{d_w}~,
\end{equation}
where the coefficients $c_{(d_z,d_w)}$ correspond to the multiplicities of the fields in the associated gauged linear $\sigma$-model (GLSM) obtained by imposing F- and D-flatness of the gauge theory. This relation connects dimer models and toric geometries in a very explicit fashion. In particular, it implies that the toric phase captured by a given dimer can be identified rather straightforwardly by computing $\operatorname{det}K$, and that two different phases can be distinguished by their toric diagrams being decorated by different sets of multiplicities. Nevertheless, it is important to emphasise that the map between toric phases and toric diagram multiplicities is strictly surjective. Due to this degeneracy, by enumerating the different sets of multiplicities that can arise in a given toric diagram, one can at most place a lower bound on the number of toric phases.

For instance, the brane tiling for the conifold in fig.~\ref{fig:conifold-dimer} produces the following Kasteleyn determinant,
\begin{align}
    P(z,w)=1-z-w-zw\quad\operatorname{mod}\quad\operatorname{SL}(3;\mathbb{Z})~,
\end{align}
which gives a square toric diagram
$\vcenter{\hbox{\begin{tikzpicture}
    \coordinate (a1) at (0,0);
    \coordinate (a2) at (0.3,0);
    \coordinate (a3) at (0,0.3);
    \coordinate (a4) at (0.3,0.3);
    \filldraw[fill=black] (a1) circle (1pt);
    \filldraw[fill=black] (a2) circle (1pt);
    \filldraw[fill=black] (a3) circle (1pt);
    \filldraw[fill=black] (a4) circle (1pt);
    \draw (a1) -- (a2) -- (a4) -- (a3) -- cycle;
\end{tikzpicture}}}\ \operatorname{mod}\ \operatorname{SL}(3;\mathbb{Z})$
with unit multiplicities.

\subsection{\texorpdfstring{$\mathbb{Z}_m\times\mathbb{Z}_n$ orbifolds of the conifold and $Y^{p,q}$ theories}{Z m x Z n orbifolds of the conifold and Y p,q theories}}

It is a very well-established notion that strings admit sensible propagation on singular backgrounds. In our investigation, in particular, we will be concerned with a certain family of hyperquotient singularities. Given the conifold $\mathcal{C}$  described in section~\ref{sec:reviewbft}, one can consider its orbifold $\mathcal{C}/\Gamma$ \cite{Bershadsky:1995sp,Aganagic:1999fe}, where $\Gamma=\mathbb{Z}_m\times\mathbb{Z}_n$ with generators
\begin{subequations}
\begin{align}
&\mathbb{Z}_m:& &(x,y,z,w)\longmapsto\left(x,y,e^{-2\pi i/m}z,e^{2\pi i/m}w\right),&\\
&\mathbb{Z}_n:& &(x,y,z,w)\longmapsto\left(e^{-2\pi i/n}x,e^{2\pi i/n}y,z,w\right),&
\end{align}
\end{subequations}
or equivalently, in terms of the bifundamental fields of the theory,
\begin{subequations}
\begin{align}
&\mathbb{Z}_m:& &(A_1,A_2,B_1,B_2)\longmapsto\left(e^{-2\pi i/m}A_1,A_2,e^{2\pi i/m}B_1,B_2\right),&\\
&\mathbb{Z}_n:& &(A_1,A_2,B_1,B_2)\longmapsto\left(e^{-2\pi i/n}A_1,A_2,B_1,e^{2\pi i/n}B_2\right).&
\end{align}
\end{subequations}
This orbifold action admits a simple interpretation when viewing the horizon $T^{1,1}$ of the conifold as a circle bundle over $S^2\times S^2$. Namely, the action above rotates both 2-spheres, while fixing the products of the poles. The action is not free on the circle fibre over these points; therefore, one finds $A_{m-1}$ and $A_{n-1}$ singularities \cite{Oh:2000ez}. As usual, when considering string propagation on such orbifolded backgrounds, there are massless fields appearing at these singular points.

Our motivation for focussing our attention to this family of geometries is multifaceted. Firstly, the orbifolded conifolds enjoy a simple description in terms of dimer models: their brane tilings correspond to square lattices with $2mn$ unit cells, while their toric diagrams are simply $(m+1)\times(n+1)$ rectangles. This feature makes the reconstruction of the corresponding Kasteleyn matrices, as well as the action of Seiberg duality, particularly straightforward. This apparent simplicity is counterpoised by a rich variety of features that recur in more complicated singularities as well. For instance, one can find and characterise a complicated structure of toric phases of the orbifolded conifold. Indeed, one can produce virtually any toric singularity by partial resolutions of such spaces. 

Furthermore, the existence of an infinite family of singularities carrying a natural (2-)parametrisation makes such geometries particularly amenable to a machine learning investigation. Indeed, the infinite nature of the family allows one to generate a sufficiently large dataset against which a neural network could be successfully trained. Additionally, the existence of the two parameters $m$ and $n$ provides two natural discrete classes for a classification model to predict. All of these points in our view build a case for orbifolds of the conifold being ideal candidates for a ``proof of principle'' work such as the present one, and lay the foundations for more generic explorations of singular geometries.

\begin{figure}
    \centering
    \includegraphics[width=\linewidth]{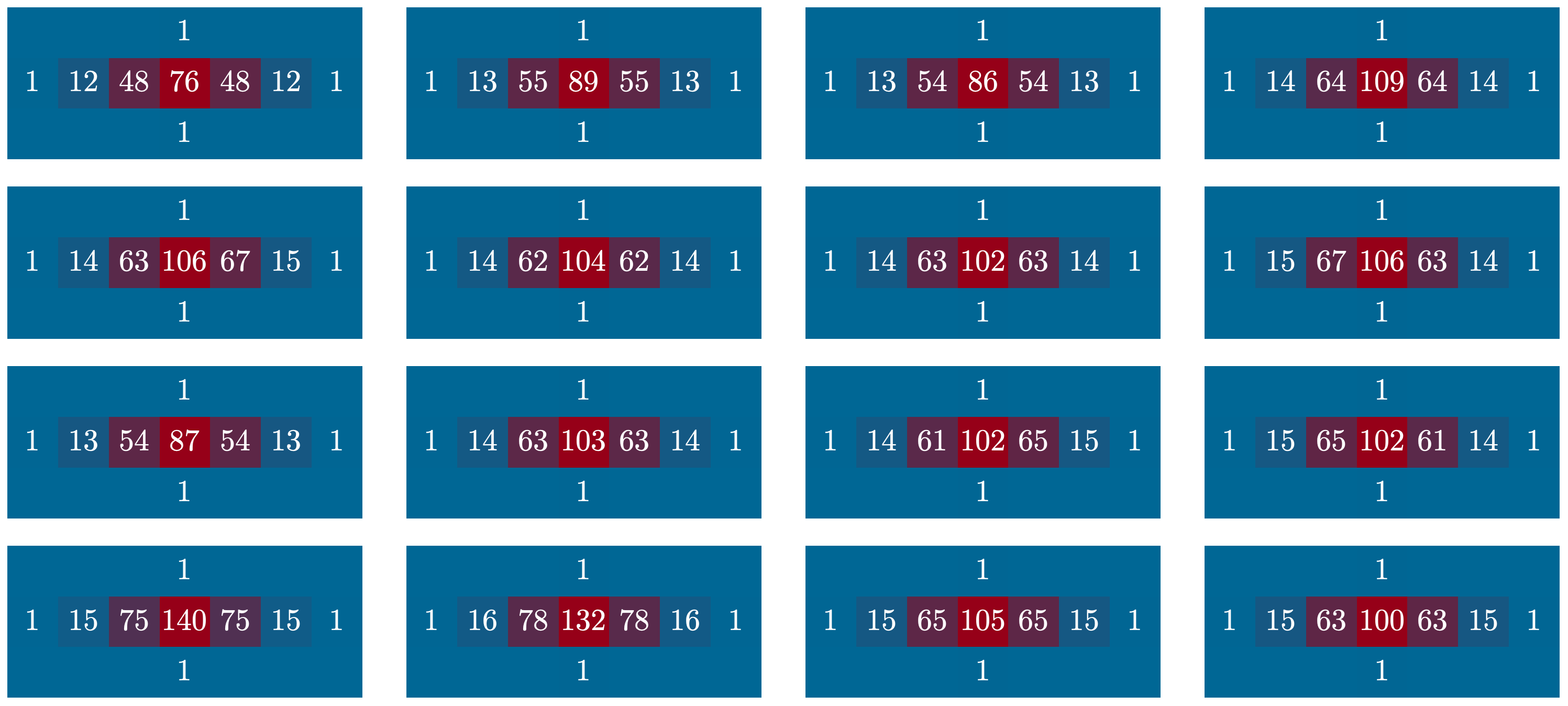}
    \caption{The GLSM multiplicities of a selection of toric phases of $Y^{6,0}$.}
    \label{fig:y60phases}
\end{figure}

For the toric phase investigation, which we will present in section~\ref{sec:results}, we will instead be concerned with the countably infinite family of $Y^{p,q}$ total spaces of $U(1)$-bundles over $S^2\times S^2$ \cite{Gauntlett:2004zh,Gauntlett:2004yd}, which -- as a consequence of Smale's classification of 5-folds \cite{e030ec85-cc49-3608-8291-834e829d7881} -- have $S^2\times S^3$ topology. Historically, those on $Y^{p,q}$ spaces were the first Sasaki-Einstein metrics to be discovered in five dimensions after the round one on $S^5$ (and $\mathbb{Z}_3$ orbifolds thereof) and the aforementioned homogeneous $T^{1,1}$ on $S^2\times S^3$. Later, it was discovered that these toric singularities are actually part of the larger infinite family $L^{a,b,c}$ \cite{Cvetic:2005ft,Martelli:2005wy,Franco:2005sm}, specified by the relation $Y^{p,q}=L^{p+q,p-q,p}$. In particular, we will focus our efforts on $Y^{6,0}$. This toric singularity can be obtained by inserting six impurities into the brane tiling of $Y^{6,6}$ \cite{Benvenuti:2004dy}, whose fundamental cell contains twelve hexagons organised into two strips. The resulting tiling consists of two rows of six squares each, and its associated toric diagram is given by the convex hull over $\mathbb{Z}^2$ of the points at $(0,0)$, $(1,0)$, $(6,6)$, and $(5,6)$ \cite{Martelli:2004wu}, modulo affine action. The $Y^{6,0}$ singularity is known to admit 18 toric phases \cite{Hanany:2005hq}, with a non-vanishing degeneracy in their multiplicity data \cite{Franco:2005fd,Hanany:2005ve}. The multiplicities of the phases we will study in section~\ref{sec:results} are shown in figure~\ref{fig:y60phases}.

\section{Investigations and results}\label{sec:results0}

In this section we present, in a systematic way, an investigation consisting in a NN-based classification of the dimers corresponding to the $\mathbb{Z}_m\times\mathbb{Z}_n$ orbifolds of the conifold and the toric phases of $Y^{6,0}$, described in the previous section. The main objective of this investigation is to determine whether standard neural networks can \emph{learn} Seiberg duality. That is to say, using standard architectures, can we efficiently classify theories of $\mathbb{Z}_m\times\mathbb{Z}_n$ orbifolds of the conifold modulo Seiberg duality, and toric phases of the $Y^{6,0}$ singularity?

Using the Kasteleyn matrix representation of a dimer, we are able to map every such tiling to a tensor. This tensor representation is a simple embedding of the space of Kasteleyn matrices into the space of tensors with three components valued in $\mathbb{Z}$; two of the components represent the coordinates (i.e. row and column) within the original Kasteleyn matrix, while the third encodes the corresponding matrix entry. For the orbifolds of the conifold, given values of $m$ and $n$, we are able to generate the corresponding Kasteleyn matrix using a readily available Mathematica package \cite{Franco:2017jeo} and apply Seiberg duality at the level of its tensorial representation, greatly facilitating data generation. For the case of $Y^{6,0}$, we generate a large number of Kasteleyn matrices describing a variety of distinct toric phases by acting iteratively with Seiberg duality onto a single starting matrix.

Importantly, for orbifolds of the conifold, we restrict ourselves to one main architecture type, namely the fully connected neural network. In our application, it is composed of an input layer, a hidden layer and an output layer, which typically contains two neurons: one for $m$ and one for $n$. As such, this network is trained in the context of a regression problem, whereby the network attempts to minimise the mean-squared error between its output and the pair $(m,n)$. Additionally, the input layer contains a custom function which reshuffles the tensor in all possible ways that preserve the underlying dimer. In order to make our investigation as broad as possible, we also perform a hyperparameter search, varying hidden layer shape, activation functions and batch sizes. All of these elements, together with other relevant notions from machine learning, are discussed in section~\ref{sec:networks}.

Furthermore, in order to evaluate the stability of our analysis against large perturbations of the dataset, we train the neural network with the above hyperparameters while purposely removing increasingly larger portions of the dataset. We will refer to such excised regions of the $(m,n)$ parameter space as ``holes''. We will also briefly consider the stability of our models against removals of specific ``depths'', i.e. matrices obtained by acting with Seiberg duality on the starting dataset a fixed number of times.

For the toric phase investigation, we initially explore a classification problem of the $Y^{6,0}$ phases using a particular kind of convolutional neural network (CNN) known as a residual network (ResNet). We then perform a finer-grained analysis, namely a regression problem using a ResNet to predict the individual GLSM multiplicities of each phase. Given the isomorphic relation between the set of labels of non-degenerate toric phases and their multiplicity content, it may at first appear that the latter investigation is simply a rephrasing of the former. While this may be true physically, the two methods crucially do differ at the machine learning level. For instance, the regression algorithm does not explicitly take the number of toric phases as an input, and so is in principle free to predict any set of multiplicities. In fact, it is also tasked with predicting the shape of the toric diagram, as it can in principle output any shape within the subset of $\mathbb{Z}^2$ it is asked to predict (in our case, this is a $7\times 3$ grid). This is to be contrasted with the classifier, which is limited to assigning probabilities over the fixed set of toric phase labels.

We start by describing in greater detail our data generation pipeline in section~\ref{sec:KasteleynGenerator}, giving a final visual representation of our dataset in figure~\ref{fig:mn_visual_holes}. section~\ref{sec:networks} is dedicated to a more thorough description of the network architectures utilised for the investigations. Notably, we describe  how the layers are connected and which hyperparameters are fine-tuned. Finally, section~\ref{sec:results} provides an account of the training results for all values of the hyperparameter search as well as the aforementioned stability analyses.

\subsection{Kasteleyn matrix generation}\label{sec:KasteleynGenerator}

\begin{figure}[t]
\centering
\begin{tikzpicture}
    \coordinate (p11) at (0,1);
    \coordinate (p12) at (0.5,1);
    \coordinate (p13) at (1,1);
    \coordinate (p21) at (0,1.5);
    \coordinate (p22) at (0.5,1.5);
    \coordinate (p23) at (1,1.5);
    \coordinate (p31) at (0,2);
    \coordinate (p32) at (0.5,2);
    \coordinate (p33) at (1,2);

    \coordinate (m11) at (5.5,1.3);

    \coordinate (m21) at (2.5,-1.9);
    \coordinate (m22) at (1.5,-1.5);
    \coordinate (m23) at (1.5,-2);

    \draw (-2,0.25) node[left]{$(m,n)$};
    \draw[->] (-1.5,0.25) -- (-0.5,0.25);


    \draw[rounded corners=0] (-0.5,0.5) -- (-0.5,2.5) -- (3.5,2.5) -- (3.5,0.5);
    \draw[fill=carmine!40!white] (-0.5,0) rectangle (3.5,0.5) node[midway]{Kasteleyn Generator};

    \fill[fill=carmine, opacity=0.8] (p11) -- (p12) -- (p22) -- (p21) -- cycle;
    \fill[fill=carmine, opacity=0.8] (p22) -- (p23) -- (p33) -- (p32) -- cycle;
    \fill[fill=MidnightBlue, opacity=0.8] (p12) -- (p13) -- (p23) -- (p22) -- cycle;
    \fill[fill=MidnightBlue, opacity=0.8] (p21) -- (p22) -- (p32) -- (p31) -- cycle;
    \draw (p11) rectangle (p33);
    \draw (p12) -- (p22) -- (p32);
    \draw (p21) -- (p22) -- (p32);
    \draw (p23) -- (p22);
    \filldraw[fill=black] (p11) circle (\nodept);
    \filldraw[fill=white] (p12) circle (\nodept);
    \filldraw[fill=black] (p13) circle (\nodept);
    \filldraw[fill=white] (p21) circle (\nodept);
    \filldraw[fill=black] (p22) circle (\nodept);
    \filldraw[fill=white] (p23) circle (\nodept);
    \filldraw[fill=black] (p31) circle (\nodept);
    \filldraw[fill=white] (p32) circle (\nodept);
    \filldraw[fill=black] (p33) circle (\nodept);

    \draw[dashed] (1.5,2.5) -- (1.5,0.5);

    \draw (2.5,1.5) node{$\begin{pmatrix}w &\cdots\\\vdots &\end{pmatrix}$};


    \draw[->] (3.5,0.25) -- (4.5,0.25);


    \draw (4.5,0) rectangle (6.9,2.5);
    \draw[fill=MidnightBlue!40!white] (4.5,0) rectangle (6.9,0.5) node[midway]{TF Tensor};

    \draw[opacity=0.125] (m11)+(0.4,0.4) node{$\begin{bmatrix}0 &\cdots\\\vdots &\end{bmatrix}$};
    \draw[opacity=0.25] (m11)+(0.2,0.2) node{$\begin{bmatrix}1 &\cdots\\\vdots &\end{bmatrix}$};
    \draw (m11) node{$\begin{bmatrix}0 &\cdots\\\vdots &\end{bmatrix}$};

    \draw[->] (5.7,0) -- (5.7,-0.5);


    \draw (4.4,-1) rectangle (7,-0.5) node[midway]{Find squares};
    \draw[->] (5.7,-1) -- (5.7,-1.5);
    \draw (4.4,-2) rectangle (7,-1.5) node[midway]{Seiberg dualise};
    \draw[->] (5.7,-2) -- (5.7,-2.5);
    \draw[thick] (5.7,-2.75) node{\huge{$\oplus$}};
    \draw[->] (5.95,-2.75) -- (7.25,-2.75) -- (7.25,0.25) -- (6.9,0.25);
    \draw (5.7,-2.9) node[below]{append/repeat};

    \draw[->] (5.45,-2.75) -- (3.5,-2.75);


    \draw (-0.5,-2.5) rectangle (3.5,-1);
    \draw[fill=yellow!40!white] (-0.5,-3) rectangle (3.5,-2.5) node[midway]{Dataset};

    \draw[opacity=0.125] (m21)+(0.3,0.3) node{\scalebox{0.75}{$\begin{bmatrix}0 &\cdots\\\vdots &\end{bmatrix}$}};
    \draw[opacity=0.25] (m21)+(0.15,0.15) node{\scalebox{0.75}{$\begin{bmatrix}1 &\cdots\\\vdots &\end{bmatrix}$}};
    \draw (m21) node{\scalebox{0.75}{$\begin{bmatrix}0 &\cdots\\\vdots &\end{bmatrix}$}};

    \draw[opacity=0.125] (m22)+(0.14,0.14) node{\scalebox{0.3}{$\begin{bmatrix}1 &\cdots\\\vdots &\end{bmatrix}$}};
    \draw[opacity=0.25] (m22)+(0.07,0.07) node{\scalebox{0.3}{$\begin{bmatrix}0 &\cdots\\\vdots &\end{bmatrix}$}};
    \draw (m22) node{\scalebox{0.3}{$\begin{bmatrix}0 &\cdots\\\vdots &\end{bmatrix}$}};

    \draw[opacity=0.125] (m23)+(0.14,0.14) node{\scalebox{0.3}{$\begin{bmatrix}0 &\cdots\\\vdots &\end{bmatrix}$}};
    \draw[opacity=0.25] (m23)+(0.07,0.07) node{\scalebox{0.3}{$\begin{bmatrix}0 &\cdots\\\vdots &\end{bmatrix}$}};
    \draw (m23) node{\scalebox{0.3}{$\begin{bmatrix}1 &\cdots\\\vdots &\end{bmatrix}$}};

    \draw (0.5,-1.75) node{$\cdots$};
    \draw (8.8,0) node{};
\end{tikzpicture}
\caption{Visual representation of our dataset generation, starting from the values of $m$ and $n$ and ending with a set of tensors, one for each Kasteleyn matrix produced via Seiberg duality.}
\label{fig:datapipeline}
\end{figure}
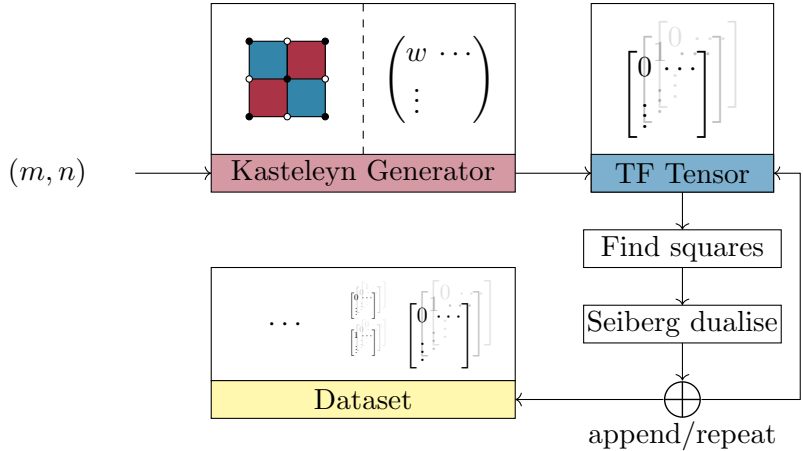

As we are utilising supervised learning methods, our results will be greatly dependent upon the size and quality of the training set. In order to construct a sufficiently large dataset in a reasonable amount of time, we must further ensure that our data generation pipeline is satisfactorily optimised. In this subsection, let us further expand upon how this was done in the context of the $\mathbb{Z}_m\times\mathbb{Z}_n$ orbifolds of the conifold. Note that scripts for all methods described in this section will soon be made available on the project's repository.

Figure~\ref{fig:datapipeline} serves as a diagrammatic representation of the data generation pipeline. The busy reader is invited to use it as a brief summary of the contents of the present section. We will aim to explain each of its blocks, precisely describing every algorithm and its corresponding input and output formats.

The first step in constructing the dataset is to generate one Kasteleyn matrix for each $(m,n)$ pair. This is done by leveraging a Mathematica package \cite{Franco:2017jeo} which constructs the brane tiling and the associated Kasteleyn matrix for $\mathbb{Z}_m\times\mathbb{Z}_n$ orbifolds of the conifold. From there, we import the matrix into Python using the SymPy package's symbolic tools. Note that while it would be straightforward to apply Seiberg duality symbolically on the Kasteleyn matrix, following \cite{Franco:2005rj}, this would not be computationally efficient. Instead, we act with Seiberg duality on a purely numerical representation of the symbolic matrix, the construction of which we shall now describe.

Converting the symbolic matrix into a numerical tensor requires one to choose a particular embedding from the space of Kasteleyn matrices into a vector space of choice. In order to preserve the matrix structure, and to further allow us to apply Seiberg duality in a straightforward manner, we choose to decompose each matrix entry into a basis of monomials represented by orthonormal vectors. The particular form of the Seiberg dualities considered in this paper implies that this orthonormal basis has finite cardinality. Let us denote by $\gamma_{w,z}$ the two cycles in the fundamental domain of the torus on which the dimer sits. Then its Kasteleyn matrix will contain polynomials of the form $\sum_{a,b\in[-1,0,1]}c_{a,b}w^az^b$. We then see that there are only nine different monomials, up to numerical coefficients. Consequently, our embedding mimics this property by setting
\begin{align}\allowdisplaybreaks
    &1 &\longrightarrow& &(1,0,0,0,0,0,0,0,0),\nonumber\\
    &w &\longrightarrow& &(0,1,0,0,0,0,0,0,0),\nonumber\\
    &1/w &\longrightarrow& &(0,0,1,0,0,0,0,0,0),\nonumber\\
    &z &\longrightarrow& &(0,0,0,1,0,0,0,0,0),\nonumber\\
    &1/z &\longrightarrow& &(0,0,0,0,1,0,0,0,0),\\
    &zw &\longrightarrow& &(0,0,0,0,0,1,0,0,0),\nonumber\\
    &z/w &\longrightarrow& &(0,0,0,0,0,0,1,0,0),\nonumber\\
    &w/z &\longrightarrow& &(0,0,0,0,0,0,0,1,0),\nonumber\\
    &1/(wz) &\longrightarrow& &(0,0,0,0,0,0,0,0,1),\nonumber
\end{align}
in the embedding map. The resulting array is a sparsely populated tensor with shape $((m+1)(n+1),(m+1)(n+1),9)$, which we readily convert into a TensorFlow SparseTensor using the \texttt{tf.sparse.from\char`_dense} method. This latter step allows us to drastically reduce the dataset size later on, while preserving computational rapidity.

\begin{figure}[t]
\centering
\begin{tikzpicture}
    \coordinate (p1) at (-1,1);
    \coordinate (p2) at (1,1);
    \coordinate (p3) at (-1,-1);
    \coordinate (p4) at (1,-1);

    \node at (0,-2) {$\downarrow$};
    \node at (5,-2) {$\downarrow$};

    \coordinate (p5) at (-1,-3);
    \coordinate (p6) at (1,-3);
    \coordinate (p7) at (-1,-5);
    \coordinate (p8) at (1,-5);
    \coordinate (p51) at (-0.5,-3.5);
    \coordinate (p61) at (0.5,-3.5);
    \coordinate (p71) at (-0.5,-4.5);
    \coordinate (p81) at (0.5,-4.5);

    \draw (p1)+(0,0.5) -- (p1) -- (p2) -- (p4) -- (p3) -- (p1);
    \draw (p1)+(-0.5,0) -- (p1);
    \draw (p2)+(0,0.5) -- (p2);
    \draw (p2)+(0.5,0) -- (p2);
    \draw (p3)+(0,-0.5) -- (p3);
    \draw (p3)+(-0.5,0) -- (p3);
    \draw (p4)+(0,-0.5) -- (p4);
    \draw (p4)+(0.5,0) -- (p4);

    \filldraw[fill=black] (p1) circle (2*\nodept);
    \filldraw[fill=white] (p2) circle (2*\nodept);
    \filldraw[fill=white] (p3) circle (2*\nodept);
    \filldraw[fill=black] (p4) circle (2*\nodept);

    \draw (p51) -- (p61) -- (p81) -- (p71) -- cycle;
    \draw (p5)+(0,0.5) -- (p5) -- (p51);
    \draw (p5)+(-0.5,0) -- (p5);
    \draw (p6)+(0,0.5) -- (p6) -- (p61);
    \draw (p6)+(0.5,0) -- (p6);
    \draw (p7)+(0,-0.5) -- (p7) -- (p71);
    \draw (p7)+(-0.5,0) -- (p7);
    \draw (p8)+(0,-0.5) -- (p8) -- (p81);
    \draw (p8)+(0.5,0) -- (p8);

    \filldraw[fill=black] (p5) circle (2*\nodept);
    \filldraw[fill=white] (p6) circle (2*\nodept);
    \filldraw[fill=white] (p7) circle (2*\nodept);
    \filldraw[fill=black] (p8) circle (2*\nodept);
    \filldraw[fill=white] (p51) circle (2*\nodept);
    \filldraw[fill=black] (p61) circle (2*\nodept);
    \filldraw[fill=black] (p71) circle (2*\nodept);
    \filldraw[fill=white] (p81) circle (2*\nodept);

    \draw (5,0) node{$\left(\begin{array}{c|c}
        A_{N\times N} & B_{N\times 2}\\
        \hline
        C_{2\times N} & \begin{matrix}
            a & b\\
            c & d
        \end{matrix}
    \end{array}\right)$};
    \draw (5,-4) node{$\left(\begin{array}{c|c|c}
        A_{N\times N} & B_{N\times 2} & 0_{N\times 2}\\
        \hline
        C_{2\times N} & 0_{2\times 2} & \begin{matrix}
            1 &\quad 0\\
            0 &\quad 1
        \end{matrix}\\
        \hline
        0_{2\times N} & \begin{matrix}
            0 & -1\\
            -1 & 0
        \end{matrix} & \begin{matrix}
        b & d\\
        a & c
        \end{matrix}
    \end{array}\right)$};
\end{tikzpicture}
\caption{Seiberg duality applied in the context of $\mathbb{Z}_m\times\mathbb{Z}_m$ orbifolds of the conifold. On the left, the effect of the duality on a quadrilateral face within the brane tiling, whose edges do not intersect the fundamental cycles $\gamma_{w,z}$ of the torus. On the right, the resultant Kasteleyn matrix transformation, where $N=(m+1)(n+1)-2$. The vanishing intersection with $\gamma_{w,z}$ implies that $a,b,c,d\in\{-1,1\}$, which greatly simplifies eq.~\ref{eq:seibergduality}.}
\label{fig:SeibergDualDimerKast}
\end{figure}
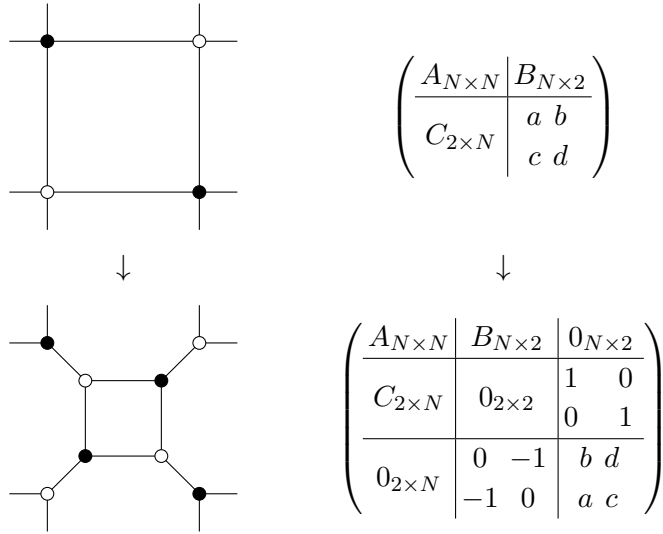

The penultimate step involves generating the Seiberg duals of the original dimer, and their numerical tensor representatives. As pointed out above, this can be done directly at the level of the sparse tensor. We refer the reader back to section~\ref{sec:reviewseiberg} for a more detailed review of the action of Seiberg duality on quadrilateral faces. For this phase of the investigation, we choose to restrict the action of Seiberg duality on quadrilateral faces that have empty intersection loci with the fundamental cycles $\gamma_{w,z}$ of the torus. In this case, the action of Seiberg duality at the level of the corresponding Kasteleyn matrix is greatly simplified, as depicted in figure~\ref{fig:SeibergDualDimerKast}.

Indeed, the simplicity of the dimers considered in this paper, and the particular Seiberg dualities we selected restrict the square matrix elements $a,b,c,d$ to take value $\pm1$ only. This bodes well for our tensor representation of the matrix and justifies our choice of a nine-component vector for each monomial, as Seiberg duality never adds any powers of $w,z$ any longer. Algorithmically, one can then search for all squares of $-1$ or $1$ within the matrix and perform the dualisation, repeat for all subsequent matrices, and once again for all values of $m$ and $n$, generating the desired dataset. Typically, we perform this loop 1000 times per pair of labels $(m,n)$.

Finally, one may notice that the dataset constructed following the steps above would yield many sparse tensors of various shapes $(M,M,9)$, with $M=(m+1)(n+1)+2d$ and $d$  being the ``depth'', i.e. the number of times Seiberg duality was applied. As our neural network architectures require a fixed input shape, we further pad every tensor to a maximal size, labelled \texttt{MATRIX\char`_SIZE}, so that our final dataset may be combined into a single sparse tensor of shape $(\texttt{DATASET\char`_SIZE},\,\texttt{MATRIX\char`_SIZE},\,\texttt{MATRIX\char`_SIZE},\,9)$. Note that padding our tensors in such a way also eliminates those on which the duality was applied too many times and whose shape exceeds the padding.

As mentioned before, for the toric phase investigation we will be interested in (a subset of) the toric phases of the $Y^{6,0}$ singularity. These can all be generated from the initial Kasteleyn matrix
\begin{align}\label{eq:y60matrix}
    \begin{pmatrix}
        -1-z & 1 & 0 & 0 & 0 & w \\
        1 & 1+\frac{1}{z} & 1 & 0 & 0 & 0 \\
        0 & 1 & -1-z & 1 & 0 & 0 \\
        0 & 0 & 1 & 1+\frac{1}{z} & 1 & 0 \\
        0 & 0 & 0 & 1 & -1-z & 1 \\
        \frac{1}{w} & 0 & 0 & 0 & 1 & 1+\frac{1}{z}
    \end{pmatrix}~.
\end{align}
In particular, we generate $5\times 10^4$ such matrices in Mathematica, reaching a maximum depth of 6. Aside from this starting point, our data generation and training pipeline is similar to that used in the previous investigation. One difference we do mention here is that in the present analysis we employ a somewhat more general algorithm to enact Seiberg duality. Indeed, in order to capture the entire breadth of the toric phase space, we act with Seiberg duality on any square face within the tiling, regardless of whether it intersects the chosen fundamental cycles of the torus or not. We do so by systematically selecting all possible combinations of four vertices in a given Kasteleyn matrix and checking whether they do form a square face (and one of minimal size, in particular) within the corresponding brane tiling. We peform this dualisation iteratively until the desired number of dual Kasteleyn matrices is achieved. In particular, we develop a routine \texttt{Seiberg\char`_Iterator} to iteratively spawn the dual matrices, which at each step calls a separate routine \texttt{Judge\char`_Seiberg} to decide whether a proposed collection of vertices within which to dualise does indeed form a fundamental square face in the brane tiling. The various toric phases which arise during this data generation process are dynamically catalogued, and a label which uniquely identifies the phase is attached to each matrix. The matrix is then converted into a vectorised form, by appending a new dimension representing each symbolic monomial, in a completely analogous fashion to the $(m,n)$ investigation. Once the data generation is complete, the Mathematica module returns the batch of matrices (in the form of SparseArray objects, which can readily be converted to TensorFlow's SparseTensor format), labels, and a dictionary to convert each label into the corresponding set of GLSM multiplicities decorating the toric diagram. Finally, this data is repackaged into a TensorFlow dataset, on which a neural network can be trained, validated, and tested.

\subsection{Network layouts}\label{sec:networks}

For our investigations, we will resort to a number of different architectures. Before delving into a detailed discussions of the network layouts, including our hyperparameter choices, let us provide a broader overview of ML methods.

An artificial \textit{neural network} is a computational model inspired by the structure and functioning of biological neurons. The primary goal of neural networks is to approximate complex, often non-linear mappings between input and output data through layers of interconnected nodes, or neurons. Mathematically, a neural network can be described as a composition of functions, where each layer represents one such function and the output of one layer becomes the input to the next.

Consider a neural network composed of $ L $ layers. Given an input vector $ x_0 \in \mathbb{R}^{n_0} $, where $n_i$ is the dimension of the $i^\text{th}$ layer, the output of the $ \ell^\text{th}$ layer is typically denoted as $ x_\ell $, and is computed recursively as
\begin{equation*}
x_\ell = \sigma(W_\ell x_{\ell-1} + b_\ell),
\end{equation*}
where $ W_\ell \in \mathbb{R}^{n_\ell \times n_{\ell-1}} $ represents the weight matrix, $ b_\ell \in \mathbb{R}^{n_\ell} $ is the bias vector, and $ \sigma $ is the \textit{activation function} applied element-wise. The function $ \sigma $ introduces non-linearity into the network, a critical feature which allows neural networks to model non-linear mappings.

The network's final layer, denoted $ x_L $, provides the model's prediction. Neural networks are typically trained in a supervised fashion, whereby the objective is to adjust the parameters $ \{W_\ell, b_\ell\}_{\ell=1}^L $ so as to minimise a predefined \textit{loss function} $ \mathcal{L}(x_L, y) $, where $ y $ represents the true target values. Optimisation is achieved through backpropagation, which computes the gradient of the loss function with respect to each parameter via the chain rule, followed by a gradient-based optimisation method such as stochastic gradient descent.

Training is often performed using \textit{mini-batch gradient descent}, whereby the dataset is divided into batches. We will also adopt this approach. If the total dataset has $ N $ samples and the chosen batch size is $ B $, the training data is partitioned into $ N/B $ batches, and the network parameters are updated after processing each batch. This strikes a balance between full-batch gradient descent and stochastic gradient descent. 

While batches and layers can be specified by a discrete number, i.e. their \textit{size}, the choice of activation function consists in picking one out of an uncountable infinity of non-linear functions. The ones that we will be focussing on are summarised below.
\begin{itemize}
    \item \textbf{Linear}: It is simply defined as 
    \[
    f(x) = x~.
    \]
    It is typically used where no non-linearity is required.

    \item \textbf{Sigmoid}: The sigmoid function maps inputs to the range $[0,1]$, and is given by 
    \[
    \sigma(x) = \frac{1}{1 + e^{-x}}~.
    \]
    It is commonly used for binary classification tasks. However, it can suffer from the vanishing gradient problem for large or small inputs.

    \item \textbf{Softmax}: The softmax function is used to normalise a vector of real numbers into a probability distribution, and is commonly applied in the output layer for multi-class classification. For a vector $z = (z_1, z_2, \dots, z_n)$, the softmax function is defined as
    \[
    \text{softmax}(z_i) = \frac{e^{z_i}}{\sum_{j=1}^{n} e^{z_j}}~,
    \]
    where $z_i$ is the $i^\text{th}$ element of the input vector.

    \item \textbf{ReLU (Rectified Linear Unit)}: The ReLU activation function is defined as 
    \[
    \text{ReLU}(x) = \max(0, x)~.
    \]
    It is widely used in hidden layers of neural networks due to its computational efficiency and ability to mitigate the vanishing gradient problem.

    \item \textbf{Leaky ReLU}: A variant of ReLU, the leaky ReLU function introduces a small slope for negative inputs, helping to address the ``dying ReLU" problem. It is defined as 
    \[
    \text{Leaky ReLU}(x) = 
    \begin{cases}
        x & \text{if } x > 0 \\
        \alpha x & \text{if } x \leq 0~.
    \end{cases}
    \]
    where $\alpha$ is a small constant, typically set to $0.01$.
\end{itemize}

Additionally to the standard types of layers which are introduced below, we constructed a custom layer, called \texttt{RandomFlip}, whose inner workings we will describe now. This layer typically follows the input layer of all our networks and serves to generalise our Kasteleyn matrix representation. Recall from our review in section~\ref{sec:reviewbft} that the Kasteleyn matrix comes about from a choice of labelling of the nodes as well as an arbitrary choice of orientations for the fundamental cycles of the torus, on which the dimer is defined. Note that there are other redundancies, such as the choice of signs assigned to various edges, which we choose not to consider here. Nevertheless, we argue that the relabelling of nodes and of the fundamental cycles, as well as changing the latter's orientation, are general enough to adequately train our networks. As an action on the matrix elements, the former amounts to permuting rows and columns while the latter corresponds to inversions and flips of $z$ and $w$. The \texttt{RandomFlip} layer, thus, serves to augment our dataset by dynamically and randomly performing these changes during training. This is akin to the possible rotations of images one may want to consider when training image classifiers.

\paragraph{Fully connected neural network}
A \textit{fully connected neural network}, also known as a \textit{dense network}, is the simplest and most commonly used type of neural network. In this type of architecture, each neuron in a given layer is connected to every neuron in the subsequent layer, hence the name. Such a structure ensures that the information from one layer is propagated entirely to the next. A fully connected neural network is completely specified by its layers and their activation functions.

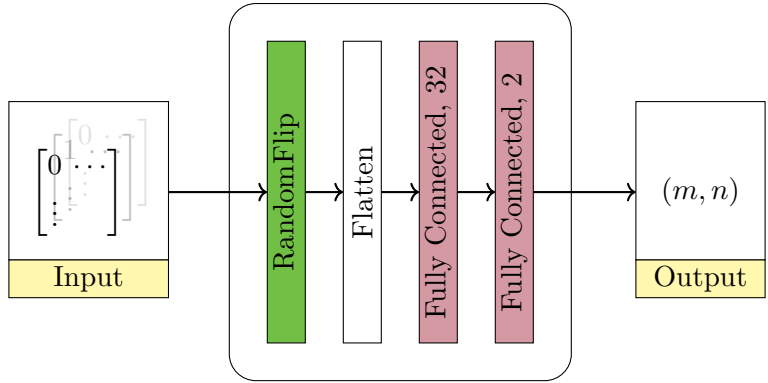
\begin{figure}[h]
    \centering
    \begin{tikzpicture}
        \coordinate (m11) at (0,0);
        \coordinate (l1) at (2.5,0);
        \coordinate (l2) at (3.5,0);
        \coordinate (l3) at (4.5,0);
        \coordinate (l4) at (5.5,0);
        \coordinate (o1) at (8.2,0);

        \draw ($(m11)+(-0.9,-0.9)$) rectangle ($(m11)+(1.2,1.2)$);
        \draw[fill=yellow!40!white] ($(m11)+(-0.9,-1.4)$) rectangle ($(m11)+(1.2,-0.9)$) node[midway]{Input};

        \draw[opacity=0.125] (m11)+(0.4,0.4) node{$\begin{bmatrix}0 &\cdots\\\vdots &\end{bmatrix}$};
        \draw[opacity=0.25] (m11)+(0.2,0.2) node{$\begin{bmatrix}1 &\cdots\\\vdots &\end{bmatrix}$};
        \draw (m11) node{$\begin{bmatrix}0 &\cdots\\\vdots &\end{bmatrix}$};

        \draw[->,thick] ($ (m11)+(1.2,0) $) -- ($(l1)$);

        \draw[fill=yellow!40!Green] ($(l1)+(0,-2)$) rectangle ($(l1)+(0.5,2)$) node[midway,rotate=90]{RandomFlip};

        \draw[->,thick] ($ (l1)+(0.5,0) $) -- ($(l2)$);

        \draw ($(l2)+(0,-2)$) rectangle ($(l2)+(0.5,2)$) node[midway,rotate=90]{Flatten};

        \draw[->,thick] ($ (l2)+(0.5,0) $) -- ($(l3)$);

        \draw[fill=carmine!40!white] ($(l3)+(0,-2)$) rectangle ($(l3)+(0.5,2)$) node[midway,rotate=90]{Fully Connected, 32};

        \draw[->,thick] ($ (l3)+(0.5,0) $) -- ($(l4)$);

        \draw[fill=carmine!40!white] ($(l4)+(0,-2)$) rectangle ($(l4)+(0.5,2)$) node[midway,rotate=90]{Fully Connected, 2};

        \draw[->,thick] ($ (l4)+(0.5,0) $) -- ($(o1)+(-0.85,0)$);

        \draw[rounded corners=10] ($(l1)+(-0.5,-2.5)$) rectangle ($(l4)+(1,2.5)$);

        \draw ($(o1)+(-0.85,-0.9)$) rectangle ($(o1)+(0.85,1.2)$);
        \draw[fill=yellow!40!white] ($(o1)+(-0.85,-1.4)$) rectangle ($(o1)+(0.85,-0.9)$) node[midway]{Output};

        \draw (o1) node{$(m,n)$};
    \end{tikzpicture}
    \caption{A diagram representing our fully connected neural network architecture. In green $\vcenter{\hbox{\begin{tikzpicture}
        \fill [yellow!40!Green] (0,0) rectangle (7pt,7pt);
    \end{tikzpicture}}}$ is our randomised input layer and in red $\vcenter{\hbox{\begin{tikzpicture}
        \fill [carmine!40!white] (0,0) rectangle (7pt,7pt);
    \end{tikzpicture}}}$ the fully connected layers, for which the activation function and the number of neurons are adjustable hyperparameters.}
    \label{fig:FullyConnected}
\end{figure}

Owing to the simplicity of the problem at hand, our fully connected network is built using only two fully connected layers, as illustrated in Figure~\ref{fig:FullyConnected}. These are preceded by our \texttt{RandomFlip} layer and a flatten layer. The latter allows every element of the input tensor to be connected to a unique neuron in the next layer. The activation functions of both fully connected layers, as well as the number of neurons in the first one, are adjustable hyperparameters. This network is trained with the Adam optimiser with constant learning rate of $0.001$, and a mean-squared error (MSE) loss function.

\paragraph{Convolutional neural network}

A \textit{convolutional neural network} (CNN) is a specialised neural network designed to handle data with grid-like topology, such as images or time-series data. In contrast to fully connected networks, CNNs exploit the local spatial structure of the data by using convolutional layers. Each convolutional layer applies a set of learnable filters (or kernels) to the input data, performing a convolution operation that slides the filter across the input and produces a feature map.

Mathematically, for an input $ x $ and a filter $ f $, the output at position $ (i,j) $ in the feature map is given by
\begin{align}
(x * f)_{i,j} = \sum_{m} \sum_{n} x_{i+m, j+n} f_{m, n} \, ,
\end{align}
where $ * $ denotes the convolution operation and the summation runs over the filter dimensions. This operation preserves the spatial relationships between input pixels, making CNNs especially effective for image-related tasks such as object recognition and segmentation.

\paragraph{Residual neural network}

A \textit{residual neural network} (ResNet) \cite{he2016deep} is a type of neural network architecture that introduces skip connections (or shortcut connections) to combat the problem of degradation in deep networks, whereby the accuracy typically worsens with depth. To illustrate this problem, consider the function $\mathcal{F}$ and a network which perfectly fits $\mathcal{F}$. One would expect that any network built from it, by adding deeper layers, should be able to fit $\mathcal{F}$ by identifying the additional layers with the identity map. However, it turns out that current solvers are unable to efficiently fit a set of non-linear layers to an identity map, making the deeper networks less performant. The skip connection bypasses this problem by adding the input of one layer to the output of a subsequent layer further down the stack. If we call $x$ such input and $\mathcal{G}$ the map realised by the stack, the skip connection essentially computes $\mathcal{G}(x)+x$. Adding such a residual block to a given network will not hinder its performance as the solver can easily fit $\mathcal{G}$ to the zero map, recovering the original network. Typically, in the context of image classification, this stack is composed of convolutional layers, whose inner workings were described previously.

This architecture enables the construction of very deep networks, such as the \textit{ResNet-50} and \textit{ResNet-101} architectures, which have 50 and 101 layers, respectively. The success of ResNet has had a profound impact on deep learning, allowing networks to be trained with hundreds or even thousands of layers.

\begin{figure}[H]
    \centering
    \begin{tikzpicture}
        \draw[fill=yellow!40!white] (-1.75,0.5) rectangle (1.75,0) node[midway]{Input};
        \draw[->,thick] (0,0) -- (0,-0.5);
        \draw[fill=yellow!40!Green] (-1.75,-0.5) rectangle (1.75,-1.0) node[midway]{RandomFlip};
        \draw[->,thick] (0,-1.0) -- (0,-1.5);
        \draw[fill=carmine!40!white] (-1.75,-1.5) rectangle (1.75,-2.0) node[midway]{7x7 Conv2D, 64};
        \draw[->,thick] (0,-2.0) -- (0,-2.5);
        \draw[fill=white] (-1.75,-2.5) rectangle (1.75,-3.0) node[midway]{Normalise/Pool};
        \draw[->,thick] (0,-3.0) -- (0,-3.5);
        \draw[fill=MidnightBlue!40!white] (-1.75,-3.5) rectangle (1.75,-4.0) node[midway]{Residual Block, 64};
        \draw[->,thick] (0,-4.0) -- (0,-4.5);
        \draw[fill=MidnightBlue!40!white] (-1.75,-4.5) rectangle (1.75,-5.0) node[midway]{Residual Block, 64};
        \draw[->,thick] (0,-5.0) -- (0,-5.5);
        \draw[fill=MidnightBlue!40!white] (-1.75,-5.5) rectangle (1.75,-6.0) node[midway]{Residual Block, 64};
        \draw[->,thick] (0,-6.0) -- (0,-6.5);
        \draw[fill=white] (-1.75,-6.5) rectangle (1.75,-7.0) node[midway]{Global Pool};
        \draw[->,thick] (0,-7.0) -- (0,-7.5);
        \draw[fill=yellow!40!white] (-1.75,-7.5) rectangle (1.75,-8.0) node[midway]{Output};

        \draw[rounded corners=10] (-2.25,1) rectangle (2.25,-8.5);
        \draw[->,thick] (5,-1) -- (5,-2);
        \draw[fill=carmine!40!white] (3.5,-2) rectangle (6.5,-2.5) node[midway]{3x3 Conv2D, 64};
        \draw[->,thick] (5,-2.5) -- (5,-3);
        \draw[fill=white] (3.5,-3) rectangle (6.5,-3.5) node[midway]{Normalise};
        \draw[->,thick] (5,-3.5) -- (5,-4);
        \draw[fill=carmine!40!white] (3.5,-4) rectangle (6.5,-4.5) node[midway]{3x3 Conv2D, 64};
        \draw[->,thick] (5,-4.5) -- (5,-5);
        \draw[fill=white] (3.5,-5) rectangle (6.5,-5.5) node[midway]{Normalise};
        \draw[->,thick] (5,-5.5) -- (5,-7);
        \draw[thick] (5,-6) node{\huge{$\oplus$}};
        
        \draw[->,thick] (5,-1.5) -- (7,-1.5) -- (7,-6) -- (5.2,-6);

        \draw[rounded corners=10] (3,-1) rectangle (7.5,-7);

        \draw[dotted,thick] (1.75,-4) -- (3.1,-6.9);
        \draw[dotted,thick] (1.75,-3.5) -- (3.1,-1.1);

    \end{tikzpicture}
    \caption{A diagram representing the ResNet architecture used in our investigation. In green $\vcenter{\hbox{\begin{tikzpicture}
        \fill [yellow!40!Green] (0,0) rectangle (7pt,7pt);
    \end{tikzpicture}}}$ is our randomized input layer, in red $\vcenter{\hbox{\begin{tikzpicture}
        \fill [carmine!40!white] (0,0) rectangle (7pt,7pt);
    \end{tikzpicture}}}$ the convolution layers, and in blue $\vcenter{\hbox{\begin{tikzpicture}
        \fill [MidnightBlue!40!white] (0,0) rectangle (7pt,7pt);
    \end{tikzpicture}}}$ the residual composite layers.}
    \label{fig:ResNet}
\end{figure}
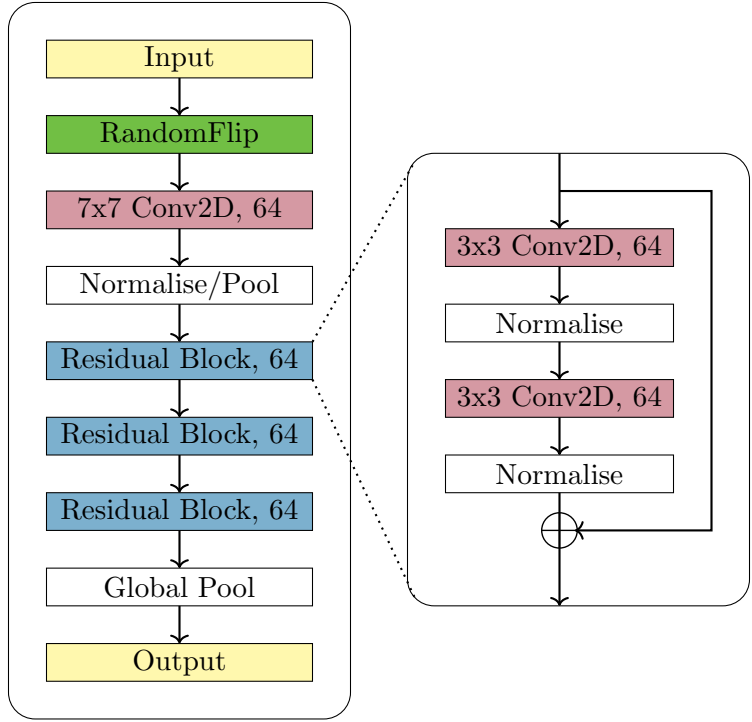

In the present case we utilise a simpler architecture, mirroring that of \textit{ResNet-50} but limited to only three residual blocks, as illustrated in figure~\ref{fig:ResNet}. As with all networks in this paper, the input layer was supplemented with our \texttt{RandomFlip} layer. Following that, are a 7x7 convolution layer with a stride of 2 and kernel size 64, a batch normalisation, a max pool layer with a pool size of 3 and a stride of 2, and three residual blocks. Finally, a global average pool precedes the output layer. In this simplified setup, every residual block is identically composed of two 3x3 convolution layer with kernel size 64, each followed by a batch normalisation. Note that each residual block does not affect the input tensor's shape, allowing for a simple application of the skip layer via tensor addition. Had the convolution layers changed the tensor shape, the standard skip layer would have also had to include a cropping/padding function to match their shapes. In all the above, ReLu was chosen as the activation function. This network is trained with the Adam optimiser with constant learning rate of $0.001$. For the classification investigation, a sparse categorical cross-entropy loss function was utilised, while a mean-squared error loss function was used for the regression problem.

\subsection{Results}\label{sec:results}

In this section, we outline the results of our investigations on the orbifolds of the conifold and the toric phases of the $Y^{6,0}$ theories. We emphasise that for each of the fully connected investigations conducted below, training $\sim 50$ epochs takes around a couple of hours on standard laptops. All of them were performed using TensorFlow. The datasets which we used for our investigations are available at \url{https://github.com/benterre/DimerML}. To the best of our knowledge, this is the only publicly available machine-learnable dataset concerning dimer models.

\paragraph{Learning $m$ and $n$ --- Precision}

\begin{figure}[t]
\centering
\includegraphics[width=\textwidth]{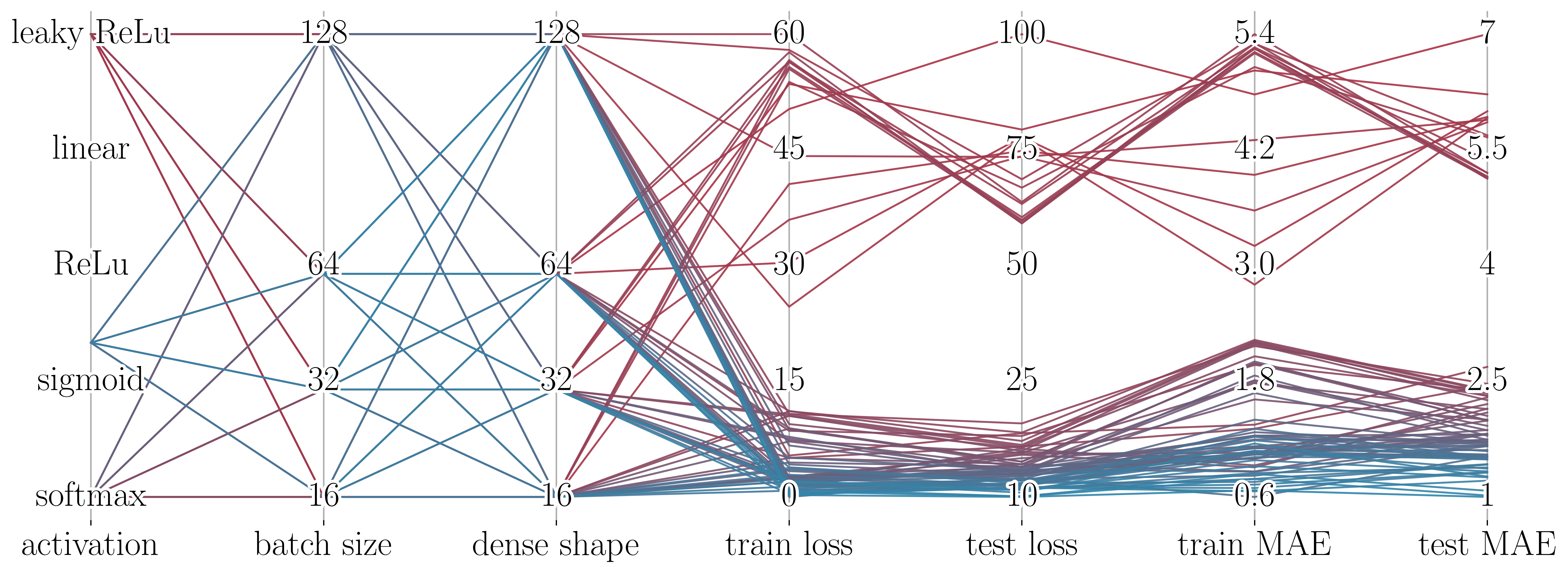}
\caption{The results of the hyperparameter search, shown with a parallel coordinates plot. Each line within the plot shows a choice of hyperparameters and its performance; both can be read by the intersection with the vertical axes. The colour reflects the performance of the network: red corresponds to poor performances while blue corresponds to good ones. The results indicate that the activation function is the hyperparameter which most influences the outcome of the investigation. Some choices, such as linear and softmax, are not suitable at all for the problem at hand. The best-performing architectures for each activation function are reported in Table \ref{tab:mnresults}.}
\label{fig:HPS}
\end{figure}

\begin{table}[t]
\centering
\begin{tabular}{|l|c|c|c|c|c|}
     \hline \diagbox[font=\footnotesize]{metric}{hparams} & \begin{tabular}{@{}c@{}}ReLU\\(64, 64)\end{tabular} & \begin{tabular}{@{}c@{}}Sigmoid\\(32, 32)\end{tabular} & \begin{tabular}{@{}c@{}}Leaky ReLU\\(64, 64)\end{tabular} & \begin{tabular}{@{}c@{}}Softmax\\(128, 128)\end{tabular} & \begin{tabular}{@{}c@{}}Linear\\(128, 64)\end{tabular} \\ \hline \hline
     R$^2$ & 0.95 & 0.94 & 0.92 & 0.85 & 0.21 \\ \hline
     MAE & 0.95 & 1.17 & 1.43 & 1.72 & 5.67 \\ \hline
\end{tabular}
\caption{The results of a subset of hyperparameter selections against R$^2$ and MAE metrics. The hyperparameters are activation function, batch size, and dense layer size. Each run consists of 50 training epochs. We show the best-performing combinations of batch and dense layer size for each activation function, whose choice appears to have a significantly greater influence on the network's performance. We select ReLu (64, 64) as the best architecture based on its performance against both metrics, and we train it for a further run of 450 epochs each; the results are reported in the main text.}
\label{tab:mnresults}
\end{table}

Let us begin by addressing the performance of the fully connected neural network in predicting the tuple $(m,n)$ for the $\mathbb{Z}_m\times\mathbb{Z}_n$ orbifolds of the conifold. As mentioned in the previous section, we test a variety of hyperparameter configurations for 50 epochs each. A configuration consists of a choice of activation function $\in\{$leaky ReLu, linear, ReLu, sigmoid, softmax$\}$, batch size $\in\{16,32,64,128\}$, and population size $\in\{16,32,64,128\}$ of neurons in the dense layer. We note that these runs are only aimed at superficially exploring the parameter space, since that comparison should be made by keeping constant the computational resources involved, and not the number of epochs (smaller batch sizes train quicker within the same number of epochs, for instance). Our results show the activation function to be the feature which most influences the architecture performance by far. The test MAE of the best-performing configurations for each activation function obtained in this preliminary investigation are collated in table~\ref{tab:mnresults}. Interestingly, the ReLu activation function consistently outperforms the others, and we find that the choices of batch size and number of neurons in the inner layer have a much smaller impact on the results than that of the activation function, although having them to match seems the most efficient choice. The broad hyperparameter scan can also be seen in figure~\ref{fig:HPS}, where the strong influence of the activation function on the final results is evident. Based on these findings, we choose to proceed with the configuration $(\text{ReLu}, 64, 64)$ and train it further. The results after 450 epochs are positive: the algorithm achieves R$^2=0.988$ and $\text{MAE}=0.722$ when evaluated against the test dataset.

\paragraph{Learning $m$ and $n$ --- Robustness}

\begin{figure}[t]
    \centering
    \begin{subfigure}[b]{0.32\textwidth}
        \centering
        \includegraphics[width=\textwidth,trim={1.1cm 1cm 1cm 1cm},clip]{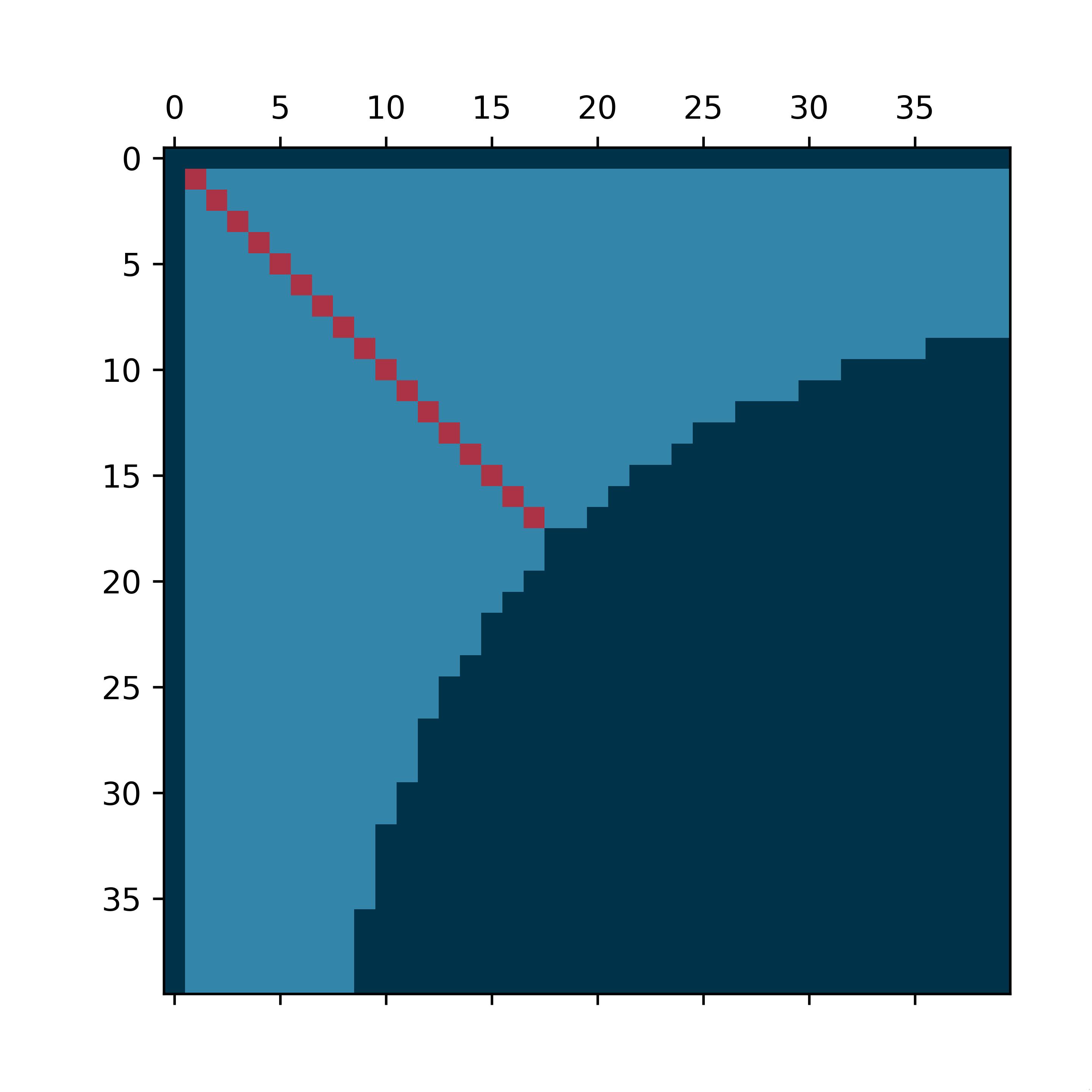}
        \caption{}
        \label{fig:mn_visual_hole_1}
    \end{subfigure}
    \hfill
    \begin{subfigure}[b]{0.32\textwidth}
        \centering
        \includegraphics[width=\textwidth,trim={1.1cm 1cm 1cm 1cm},clip]{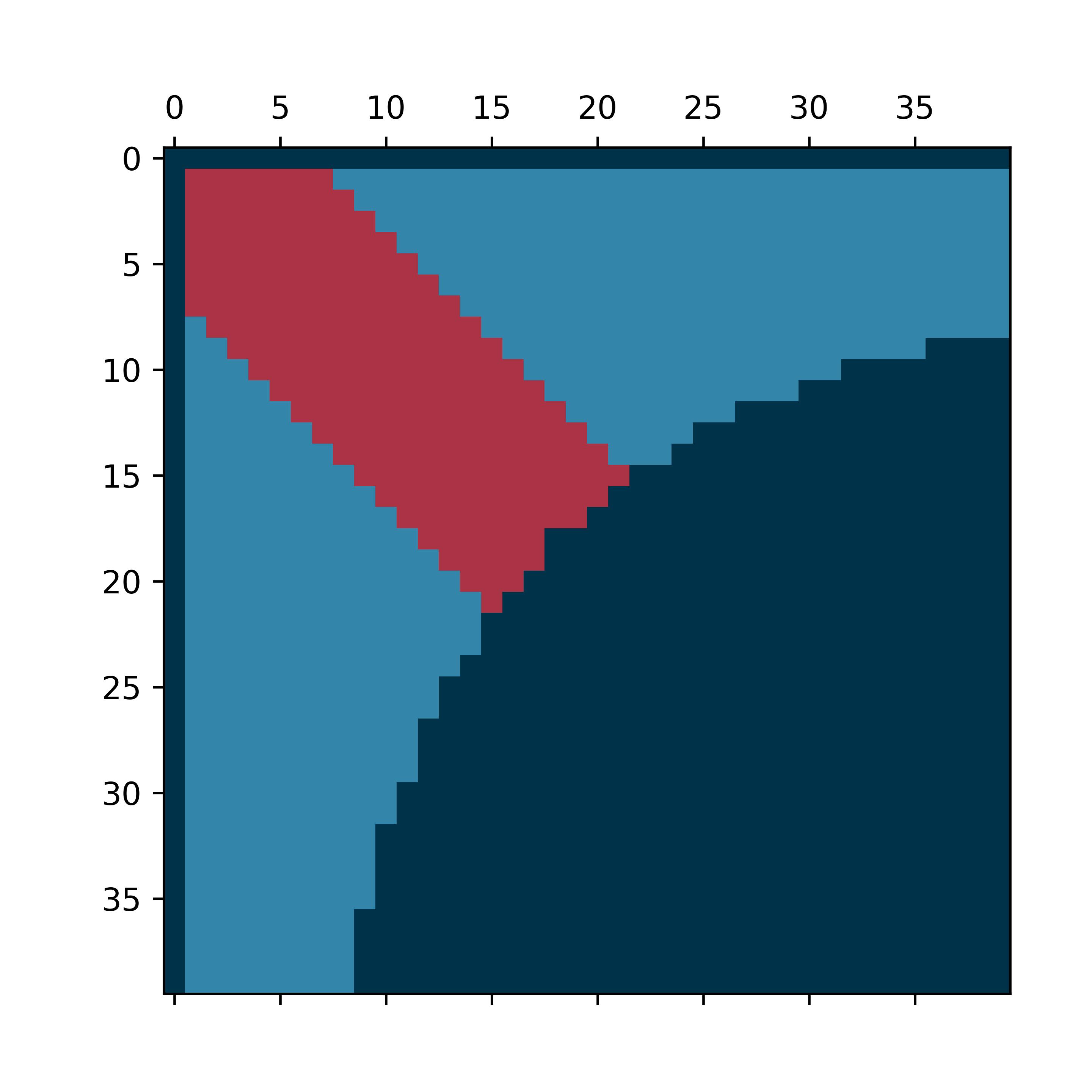}
        \caption{}
        \label{fig:mn_visual_hole_2}
    \end{subfigure}
    \hfill
    \begin{subfigure}[b]{0.32\textwidth}
        \centering
        \includegraphics[width=\textwidth,trim={1.1cm 1cm 1cm 1cm},clip]{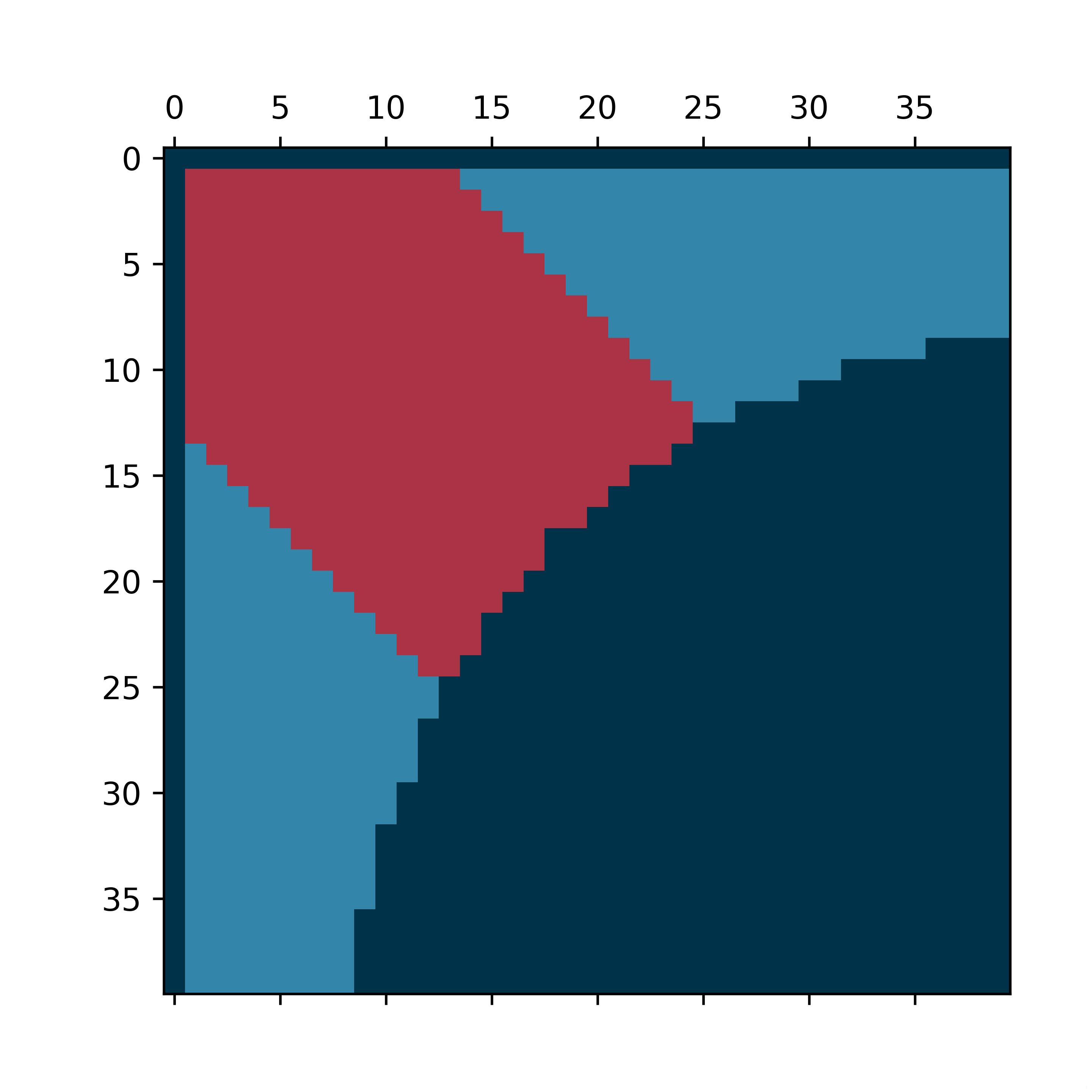}
        \caption{}
        \label{fig:mn_visual_hole_3}
    \end{subfigure}\newline
    \begin{subfigure}[b]{0.32\textwidth}
        \centering
        \includegraphics[width=\textwidth,trim={1.1cm 1cm 1cm 1cm},clip]{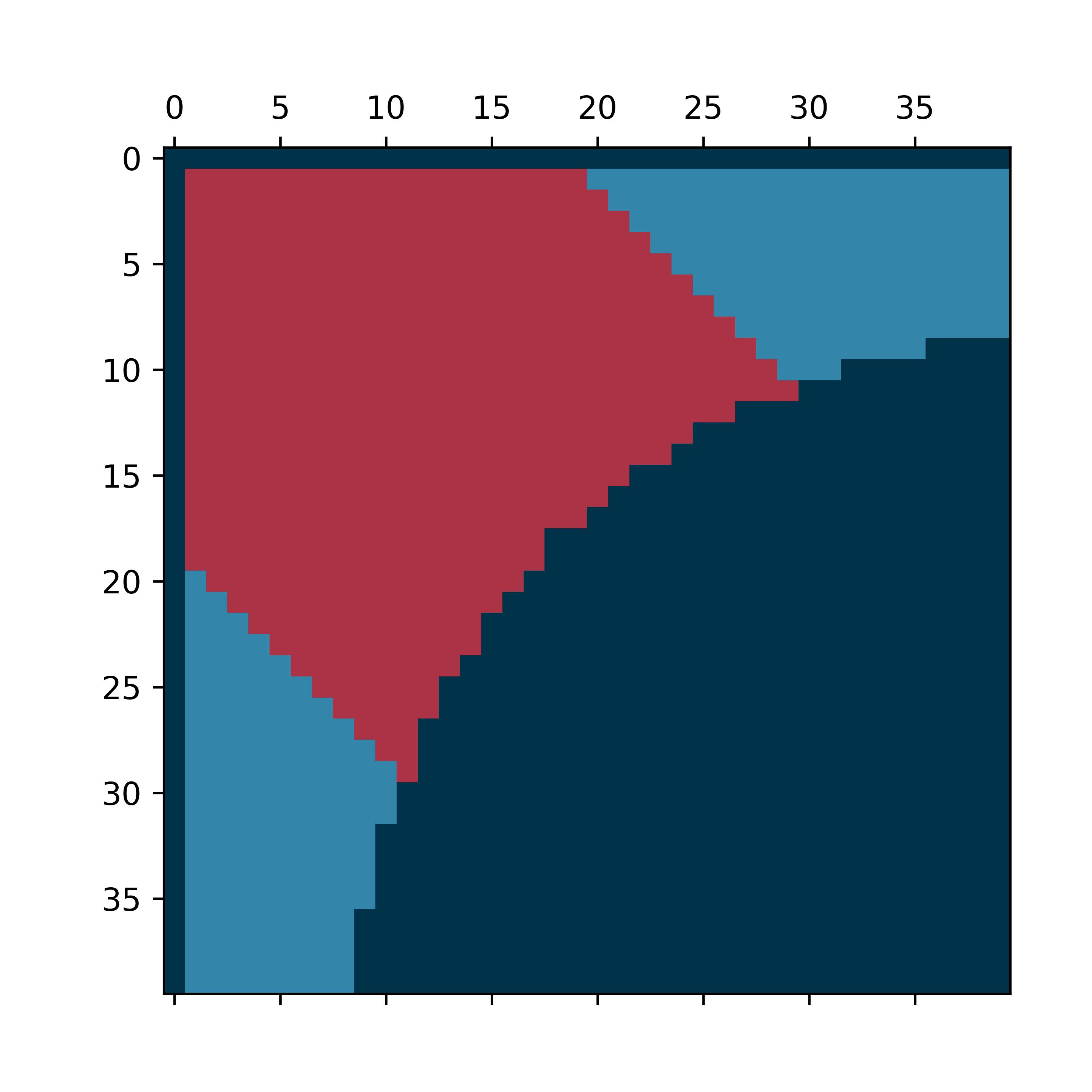}
        \caption{}
        \label{fig:mn_visual_hole_4}
    \end{subfigure}
    \hfill
    \begin{subfigure}[b]{0.32\textwidth}
        \centering
        \includegraphics[width=\textwidth,trim={1.1cm 1cm 1cm 1cm},clip]{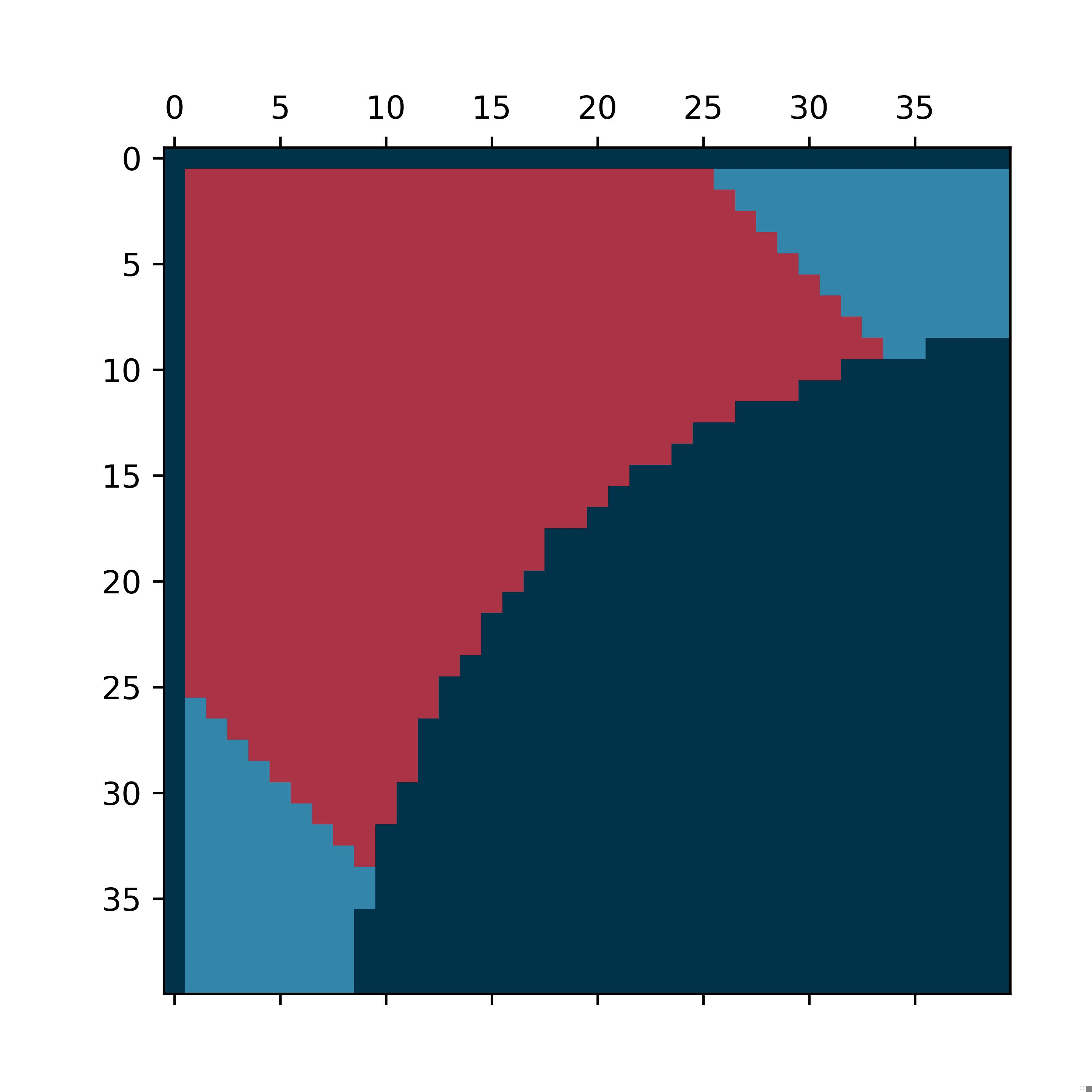}
        \caption{}
        \label{fig:mn_visual_hole_5}
    \end{subfigure}
    \hfill
    \begin{subfigure}[b]{0.32\textwidth}
        \centering
        \includegraphics[width=\textwidth,trim={1.1cm 1cm 1cm 1cm},clip]{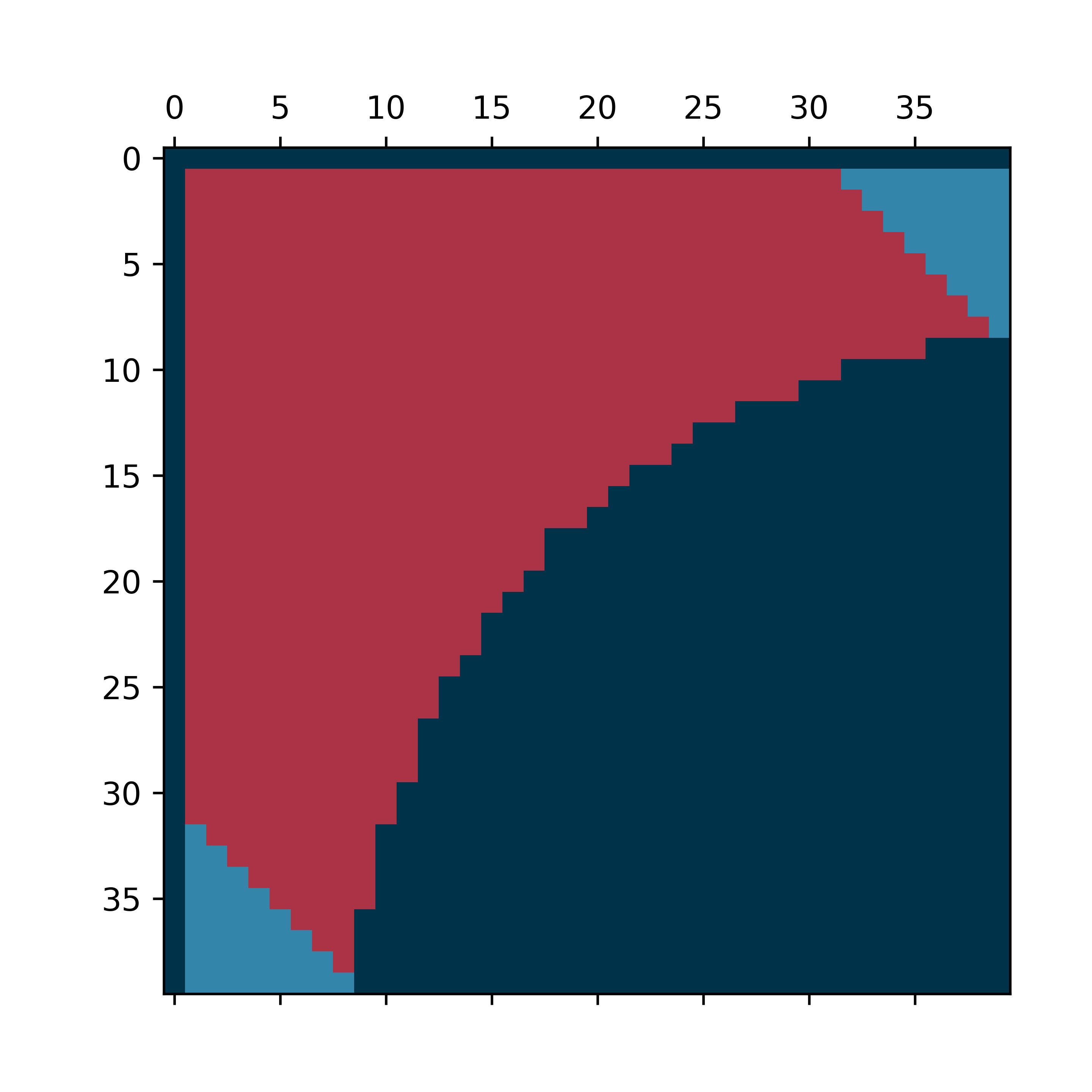}
        \caption{}
        \label{fig:mn_visual_hole_6}
    \end{subfigure}
    \caption{(a-f): A visual representation of our data selection process for hole size $=$ 1, 7, 13, 19, 25, and 31, respectively. The graphs are plotted in $(m,n)$-space. In light blue $\vcenter{\hbox{\begin{tikzpicture}
        \fill [clr1] (0,0) rectangle (7pt,7pt);
    \end{tikzpicture}}}$ are the values of $(m,n)$ on which the NN is trained, in red $\vcenter{\hbox{\begin{tikzpicture}
        \fill [clr2] (0,0) rectangle (7pt,7pt);
    \end{tikzpicture}}}$ those purposely excluded from training but included in testing (i.e. the ``hole''), and in dark blue $\vcenter{\hbox{\begin{tikzpicture}
        \fill [clr3] (0,0) rectangle (7pt,7pt);
    \end{tikzpicture}}}$ those entirely omitted (i.e. never generated nor tested on).}
    \label{fig:mn_visual_holes}
\end{figure}

Our final investigation for the $\mathbb{Z}_m\times\mathbb{Z}_n$ orbifolds of the conifold consists in evaluating the stability of our networks against the excision of portions of variable size from the training dataset. 

\begin{figure}[t]
\centering
    \begin{subfigure}[b]{0.45\textwidth}
        \centering
        \raisebox{5pt}{\includegraphics[width=\textwidth]{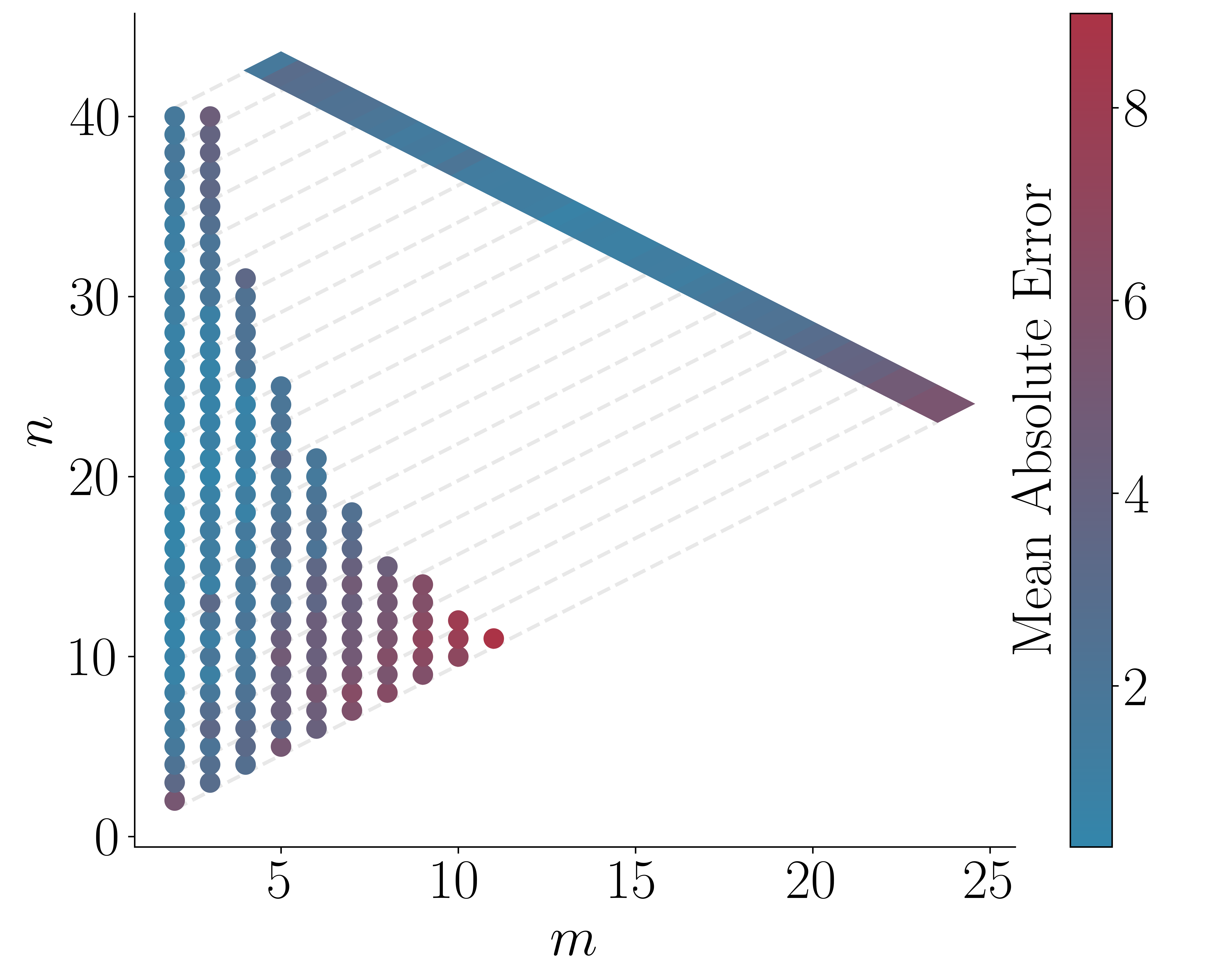}}
        \caption{}
    \end{subfigure}
    \hfill
    \begin{subfigure}[b]{0.5\textwidth}
        \centering
        \includegraphics[width=\textwidth]{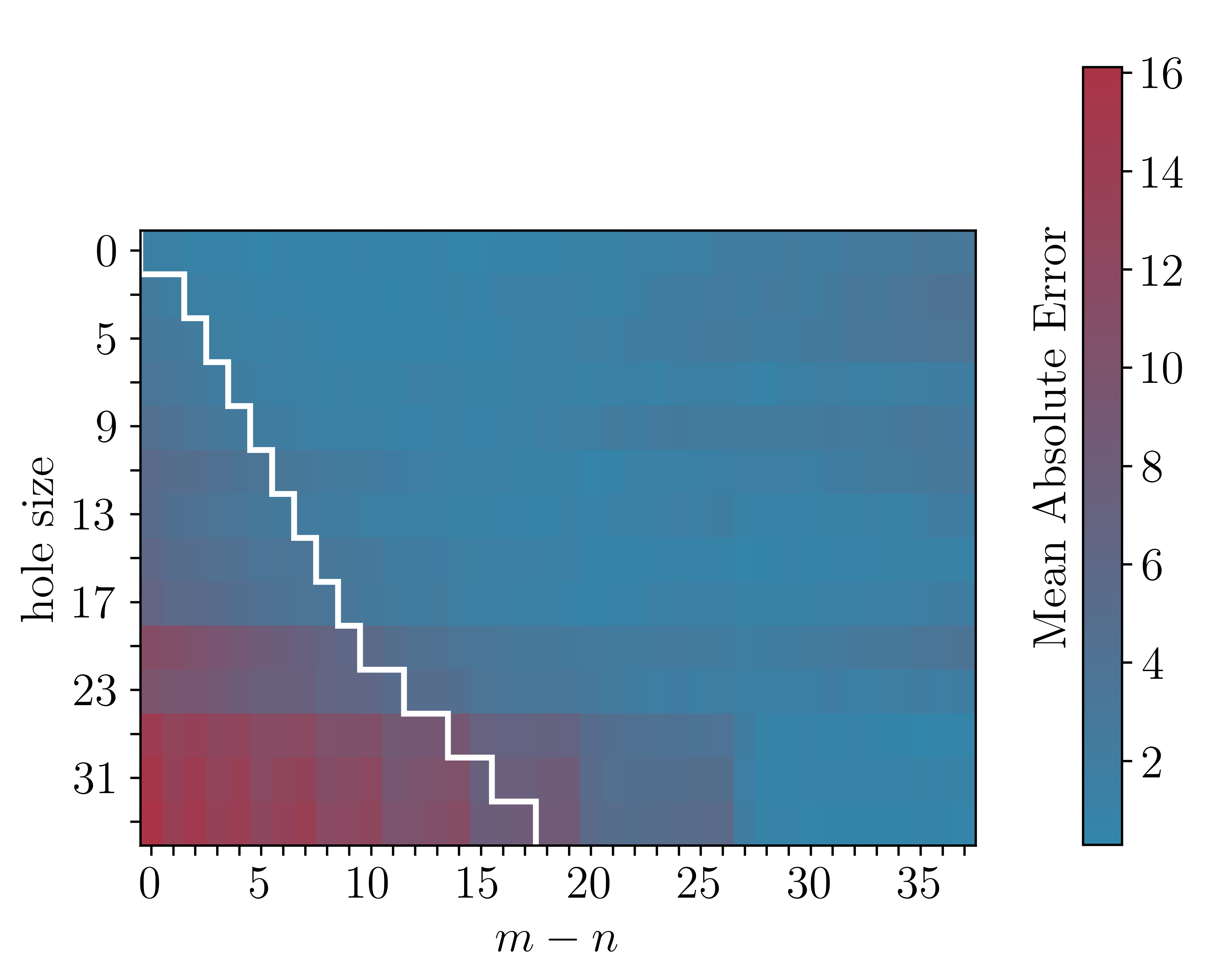}
        \caption{}
    \end{subfigure}
\caption{In this figure, we display the performance of the network for various hole sizes. In (a), we show how a 1d strip of MAE values as a function of distance from the $m=n$ diagonal is obtained from the data over the $(m,n)$ plane, namely by taking the average over the diagonals marked by dashed lines. The figure shows the results for the dataset with a hole size of 11. In (b), we plot the MAE at various distances from the hole (horizontal axis) and for various hole sizes (vertical axis). For each strip, the hole is delimited by a white outline.}
\label{fig:plot_col}
\end{figure}

As part of this investigation, we train a single architecture on datasets with larger and larger ``holes" in the $(m,n)$ plane. We refer the reader to Figure~\ref{fig:mn_visual_holes} for a visual representation of the holes we consider. We select the highest achieving architecture found previously, namely a single dense layer with 64 neurons and ReLu activation function, and a batch size of 64. In every run, the architecture is trained for 50 epochs and then tested against the full $(m,n)$ dataset, including samples from within the hole. In particular, we measure the network's performance separately for each value of $(m,n)$, as to verify whether the network is able to satisfactorily learn Seiberg duality within the hole as well as on seen values of $(m,n)$, or the extent to which it fails to do so. Note that obtaining a high precision and accuracy, even on unseen data, does not prove any \textit{learning}, in the human sense, per se. One can confidently claim that a neural network has learnt Seiberg duality only when it has regressed the rules behind this operation. Nevertheless, the ability to predict the features of Seiberg duality within a certain regime, i.e.~for a given subset of the $(m,n)$ plane, is still a remarkable result, albeit not the end of the story. The investigation for different excised datasets was performed in this spirit, and its results for various hole sizes are summarised in figure~\ref{fig:plot_col}. 

As expected, we see that the learning process of the neural network is inherently \textit{local} in nature. The information collected in the regions selected for training is not well-interpolated across the hole, unless the latter's size is reasonably small. In other words, the supervised learning captures efficiently the local patterns; a substantially larger dataset might be needed for it to be able to recognise the global structure of Seiberg duality. Since the one presented in this paper is only, to the best of our knowledge, the first publicly available instance of a machine-learnable dataset of Seiberg-dualised Kasteleyn matrices, we see ample space for improvement in this regard.

\begin{figure}[t]
\centering
\includegraphics[width=0.6\textwidth]{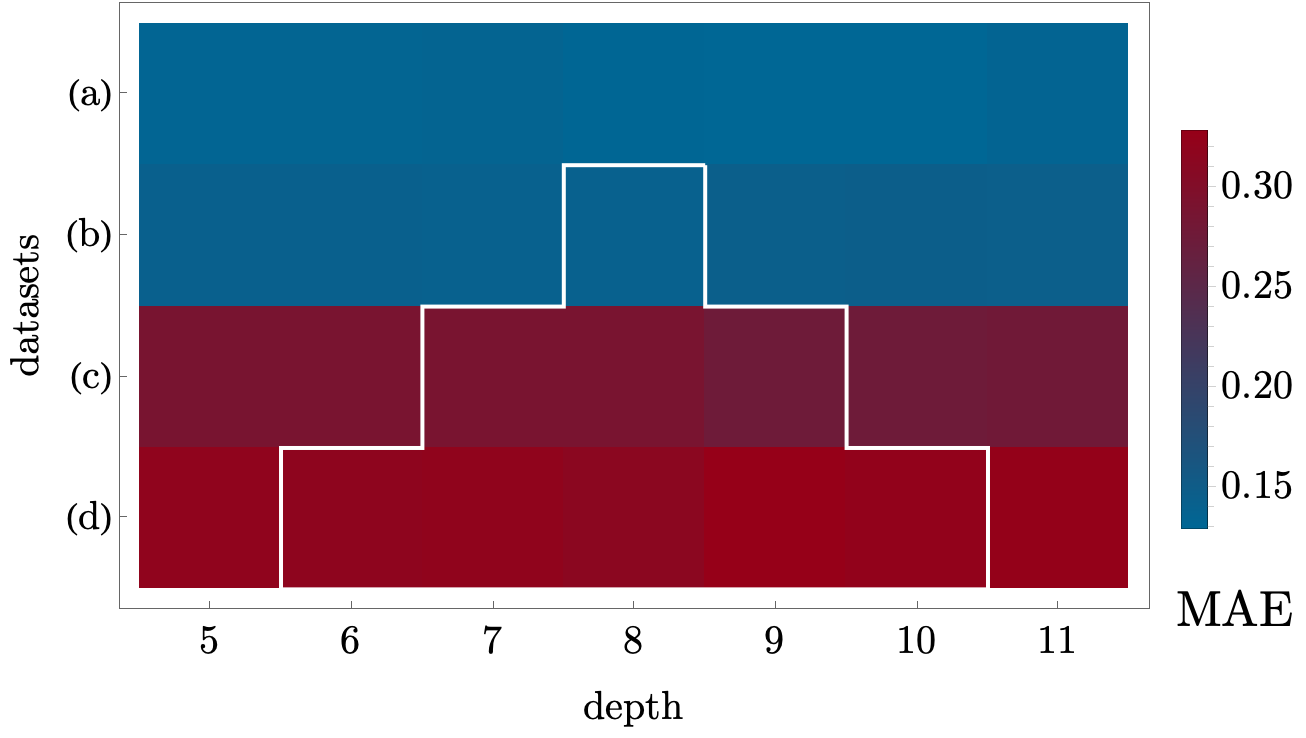}
\caption{The MAE at various depths for (a) the full dataset, and the datasets with excisions at (b) $d=8$, (c) $d\in\{7,8,9\}$, (d) $d\in\{6,7,8,9\}$, with the excised regions marked by a white outline for clarity. Larger dataset excisions affect the performance uniformly in $d$, and no hole appears. The relatively large drop in performance from datasets (b) to (c) could be caused by the excised region jumping from accounting for 14\% to 43\% of the dataset.}
\label{fig:depthholes}
\end{figure}

Finally, we perform a second robustness investigation by excising holes in depth space rather than in $(m,n)$ space. As the number of Kasteleyn matrices grows significantly with depth, we restrict this investigation to tuples $(m,n)$ with $6<mn<19$, and randomly select from those theories 800 matrices at each depth $d$ for $4<d<12$; this allows us to work with a balanced dataset. We train the same dense architecture discussed in the previous section on four datasets: the full one just described, a dataset with an excision at $d=8$, one with an excision at $d\in\{7,8,9\}$, and finally one with an excision at $d\in\{6,7,8,9,10\}$. The results are summarised in figure~\ref{fig:depthholes}. Interestingly, we find that, while an increasingly larger hole gradually decreases the performance of the network, as one might have expected, it does so uniformly in depth space, up to statistical fluctuations. In other words, we do not observe the appearance of a statistically significant ``hole'', or drop in performance, at the excised region in depth space. This is evidence that the network is able to correctly interpolate to unseen depths and has in this sense achieved an understanding of Seiberg duality which is global.

\paragraph{Learning the toric phase space of $Y^{6,0}$}

We end by reporting an investigation of a different (and more challenging) nature: predicting the toric phases of the $Y^{6,0}$ theories. This problem can be tackled in two ways. One can leverage the finite cardinality of the set of toric phases of $Y^{6,0}$ to set up a classification task, consisting in feeding the network a set of unique identifiers, each labelling a toric phase (up to degeneracies thereof). Alternatively, one can explore if the individual multiplicities can be machine learnt from scratch; this would clearly be a more ambitious investigation.

Given the non-trivial nature of this task, we find we must resort to more sophisticated architectures than the fully connected networks used previously; in particular, we settle on the ResNet architecture, described in section~\ref{sec:networks}.

\begin{table}[b]
\centering
\begin{tabular}{lllllll}
\cline{2-3} \cline{6-7}
\multicolumn{1}{l|}{} & \multicolumn{2}{c|}{\begin{tabular}{@{}c@{}}ResNet\\(classifier)\end{tabular}} & & \multicolumn{1}{l|}{} & \multicolumn{2}{c|}{\begin{tabular}{@{}c@{}}ResNet\\(regressor)\end{tabular}} \\
\cline{2-3} \cline{6-7} 
\multicolumn{1}{l|}{}  & \multicolumn{1}{c|}{w/o \texttt{RF}L} & \multicolumn{1}{c|}{w/ \texttt{RF}L} & & \multicolumn{1}{l|}{} & \multicolumn{1}{c|}{w/o \texttt{RF}L} & \multicolumn{1}{c|}{w/ \texttt{RF}L} \\ \cline{1-3} \cline{5-7} \noalign{\vskip\doublerulesep\vskip-\arrayrulewidth} \cline{1-3} \cline{5-7}
\multicolumn{1}{|l|}{epochs} & \multicolumn{1}{c|}{300} & \multicolumn{1}{c|}{30000} & \multicolumn{1}{l|}{} & \multicolumn{1}{l|}{epochs} & \multicolumn{1}{c|}{300} & \multicolumn{1}{c|}{19000} \\ \cline{1-3} \cline{5-7} 
\multicolumn{1}{|l|}{accuracy} & \multicolumn{1}{c|}{$1.0\pm0.0$} & \multicolumn{1}{c|}{0.7173} & \multicolumn{1}{l|}{} & \multicolumn{1}{l|}{R$^2$} & \multicolumn{1}{c|}{$1.0\pm 0.0$} & \multicolumn{1}{c|}{$0.9933$} \\ \cline{1-3} \cline{5-7}     
& & & \multicolumn{1}{l|}{} & \multicolumn{1}{l|}{MAE} & \multicolumn{1}{c|}{$0.021\pm0.001$} & \multicolumn{1}{c|}{$0.5853$} \\ \cline{5-7}  
\end{tabular}
\caption{Results for our classification (coarse) and regression (fine) investigations of the toric phase space of $Y^{6,0}$ using a ResNet model, with and without the \texttt{RandomFlip} layer (\texttt{RF}L). The uncertainties shown are obtained by taking the average of each model over 3 training runs.}
\label{tab:toric_results}
\end{table}

In both the classifier and regressor investigations, we find that the \texttt{RandomFlip} layer has a profound impact on the performance of the network. This is explicitly visible in the results collated in table~\ref{tab:toric_results}. For the ResNet classifier, we showcase the accuracy achieved by the network in predicting the single phase identifier of each Kasteleyn matrix. The regressor is instead tasked with predicting the individual multiplicities in the region $[0,2]\times[0,6]\subset\mathbb{Z}^2$. Therefore, within this region, the algorithm is in principle free to predict any shape for the toric diagram, even non-convex ones. We report both the R$^2$ and MAE results found in this effort.

As reported in table~\ref{tab:toric_results}, in the absence of the \texttt{RandomFlip} layer, both the classifier and the regressor are able to achieve extremely positive results against all metrics considered in a relatively limited number of epochs. On the other hand, the inclusion of the \texttt{RandomFlip} layer means that both architectures must be trained for a significantly larger number of epochs to achieve satisfactory results. This suggests that, for a fixed choice of labelling of the vertices and for a fixed choice of fundamental cycles $\gamma_{w,z}$, the network is able to classify toric phases and predict their multiplicity content to a fair extent; however, it severely suffers from relabellings performed for the purpose of augmenting the dataset. This behaviour may urge one to consider instead an architecture to which such conventions are transparent, i.e. one for which trivial invariances such as vertex relabelling are built-in. We shall return to this point briefly in the next section. We conclude this section by showcasing a set of sample predictions performed by the ResNet regressor without \texttt{RandomFlip} layer in table~\ref{tab:resnet_examples}.

\begin{table}[t]
    \centering
    \begin{tabular}{cccc}
        depth & tiling & toric data $\quad$ & predictions (3 d.p.)\\  
        0 & 
       \raisebox{-.5\height}{\includegraphics[width=0.3\textwidth]{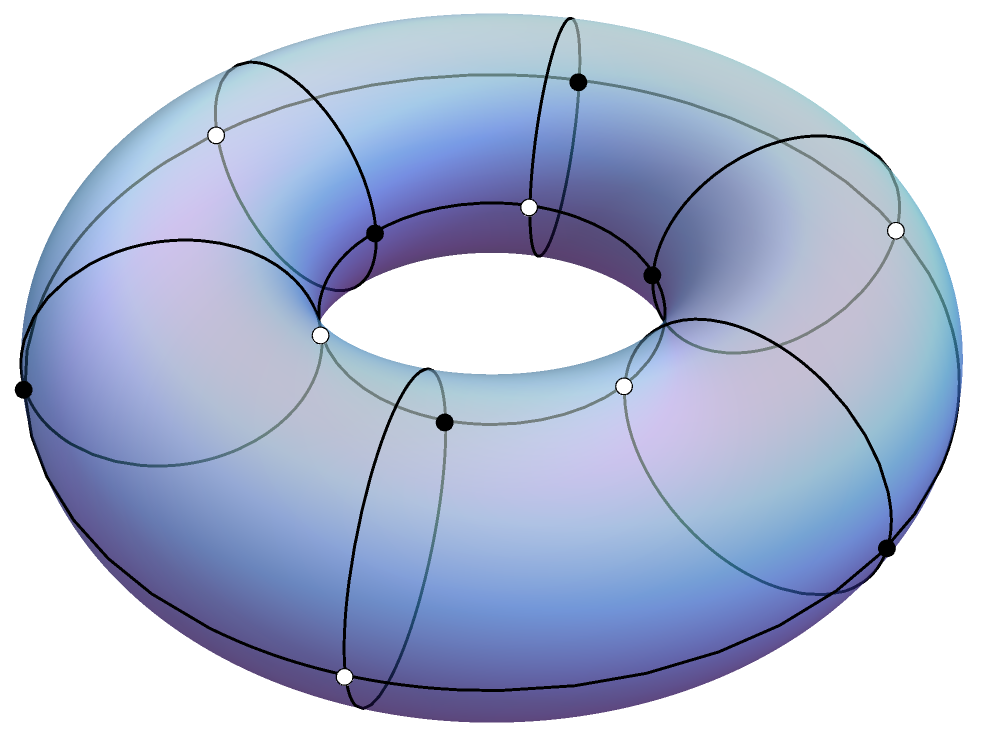}} &
       $\vcenter{\hbox{\begin{tikzpicture}
            \filldraw[fill=black] (0,0) circle (\nodept);
            \filldraw[fill=black] (0,0.5) circle (\nodept);
            \filldraw[fill=black] (0,1) circle (\nodept);
            \filldraw[fill=black] (0,1.5) circle (\nodept);
            \filldraw[fill=black] (0,2) circle (\nodept);
            \filldraw[fill=black] (0,2.5) circle (\nodept);
            \filldraw[fill=black] (0,3) circle (\nodept);
            \filldraw[fill=black] (-0.8,1.5) circle (\nodept);
            \filldraw[fill=black] (0.8,1.5) circle (\nodept);
            
            \draw[densely dotted] (0,0) -- (0.8,1.5) -- (0,3) -- (-0.8,1.5) -- cycle;
    
            \node[text=black] at (0.4,0.5) {\contour{white}{$12$}};
            \node[text=black] at (0.4,1) {\contour{white}{$48$}};
            \node[text=black] at (0.4,1.5) {\contour{white}{$76$}};
            \node[text=black] at (0.4,2) {\contour{white}{$48$}};
            \node[text=black] at (0.4,2.5) {\contour{white}{$12$}};
            \node[text=black] at (0.3,0) {\contour{white}{$1$}};
            \node[text=black] at (0.3,3) {\contour{white}{$1$}};
            \node[text=black] at (-0.5,1.5) {\contour{white}{$1$}};
            \node[text=black] at (1.1,1.5) {\contour{white}{$1$}};
            
            \node[text=black] at (1.5,2.5) {};
            
        \end{tikzpicture}}}$  &
        \begin{tabular}{ |p{0.9cm}|p{1.2cm}|p{0.9cm}| } 
         \hline
         0.003 & 1.006 & 0.000 \\ \hline
         0.003 & 11.901  & 0.003  \\ \hline
         0.002 & 47.937 & 0.002  \\ \hline
         1.005 & 75.781 & 1.005  \\ \hline
         0.001 & 48.145 & 0.002  \\ \hline
         0.008 & 11.977  & 0.003  \\ \hline
         0.000 & 1.005 & 0.003  \\ \hline
        \end{tabular}
        \\ 2 &
         \raisebox{-.5\height}{\includegraphics[width=0.3\textwidth]{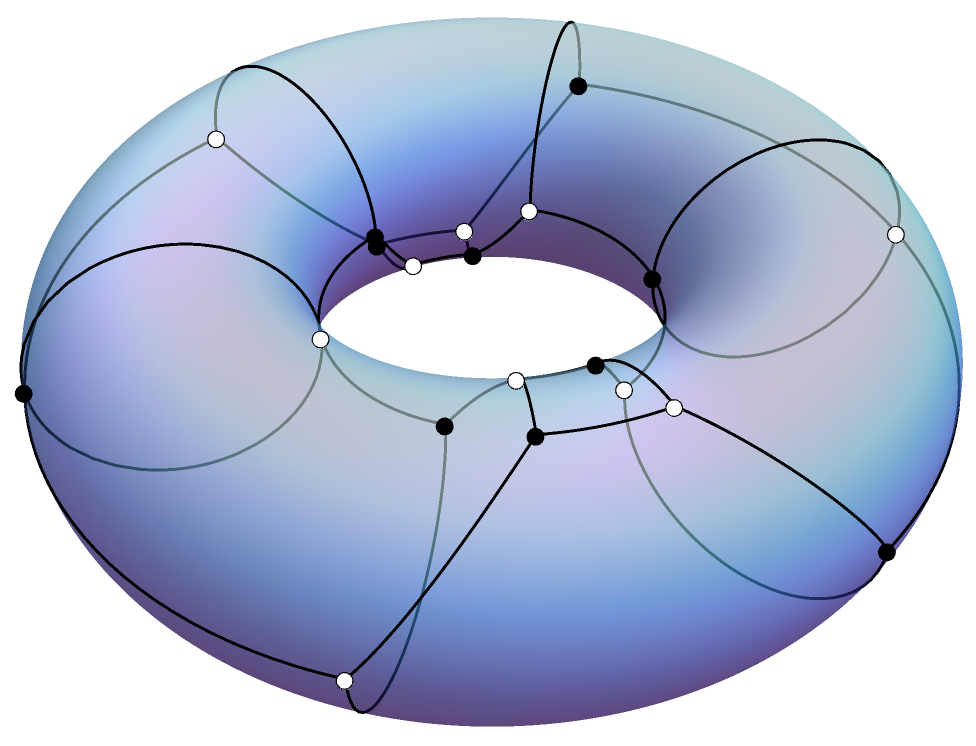}} &
       $\vcenter{\hbox{\begin{tikzpicture}
            \filldraw[fill=black] (0,0) circle (\nodept);
            \filldraw[fill=black] (0,0.5) circle (\nodept);
            \filldraw[fill=black] (0,1) circle (\nodept);
            \filldraw[fill=black] (0,1.5) circle (\nodept);
            \filldraw[fill=black] (0,2) circle (\nodept);
            \filldraw[fill=black] (0,2.5) circle (\nodept);
            \filldraw[fill=black] (0,3) circle (\nodept);
            \filldraw[fill=black] (-0.8,1.5) circle (\nodept);
            \filldraw[fill=black] (0.8,1.5) circle (\nodept);
            
            \draw[densely dotted] (0,0) -- (0.8,1.5) -- (0,3) -- (-0.8,1.5) -- cycle;
    
            \node[text=black] at (0.4,0.5) {\contour{white}{$14$}};
            \node[text=black] at (0.4,1) {\contour{white}{$63$}};
            \node[text=black] at (0.4,1.5) {\contour{white}{$102$}};
            \node[text=black] at (0.4,2) {\contour{white}{$63$}};
            \node[text=black] at (0.4,2.5) {\contour{white}{$14$}};
            \node[text=black] at (0.3,0) {\contour{white}{$1$}};
            \node[text=black] at (0.3,3) {\contour{white}{$1$}};
            \node[text=black] at (-0.5,1.5) {\contour{white}{$1$}};
            \node[text=black] at (1.1,1.5) {\contour{white}{$1$}};
            
            \node[text=black] at (1.5,2.5) {};
            
        \end{tikzpicture}}}$  &
        \begin{tabular}{ |p{0.9cm}|p{1.2cm}|p{0.9cm}| } 
         \hline
         0.004 & 1.010 & 0.000 \\ \hline
         0.004 & 13.958 & 0.004 \\ \hline
         0.002 & 62.887 & 0.002 \\ \hline
         1.009 & 102.025 & 1.010 \\ \hline
         0.001 & 62.951 & 0.003 \\ \hline
         0.010 & 13.989 & 0.004 \\ \hline
         0.000 & 1.010  & 0.005 \\ \hline
        \end{tabular}
        \\ 6 &
        \raisebox{-.5\height}{\includegraphics[width=0.3\textwidth]{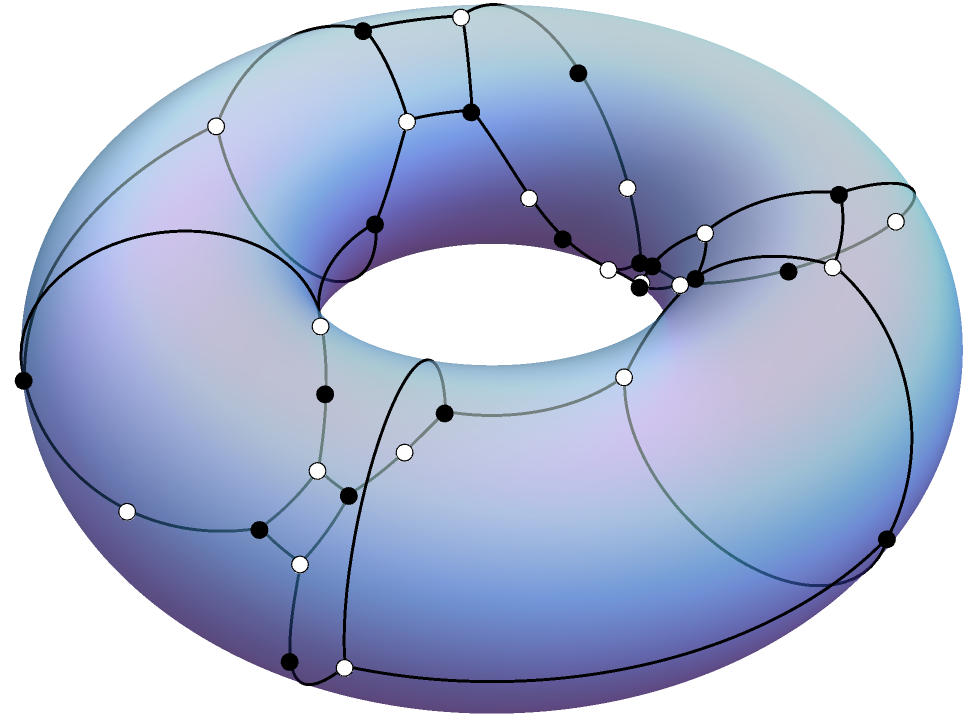}} &
       $\vcenter{\hbox{\begin{tikzpicture}
            \filldraw[fill=black] (0,0) circle (\nodept);
            \filldraw[fill=black] (0,0.5) circle (\nodept);
            \filldraw[fill=black] (0,1) circle (\nodept);
            \filldraw[fill=black] (0,1.5) circle (\nodept);
            \filldraw[fill=black] (0,2) circle (\nodept);
            \filldraw[fill=black] (0,2.5) circle (\nodept);
            \filldraw[fill=black] (0,3) circle (\nodept);
            \filldraw[fill=black] (-0.8,1.5) circle (\nodept);
            \filldraw[fill=black] (0.8,1.5) circle (\nodept);
            
            \draw[densely dotted] (0,0) -- (0.8,1.5) -- (0,3) -- (-0.8,1.5) -- cycle;
    
            \node[text=black] at (0.4,0.5) {\contour{white}{$14$}};
            \node[text=black] at (0.4,1) {\contour{white}{$64$}};
            \node[text=black] at (0.4,1.5) {\contour{white}{$109$}};
            \node[text=black] at (0.4,2) {\contour{white}{$64$}};
            \node[text=black] at (0.4,2.5) {\contour{white}{$14$}};
            \node[text=black] at (0.3,0) {\contour{white}{$1$}};
            \node[text=black] at (0.3,3) {\contour{white}{$1$}};
            \node[text=black] at (-0.5,1.5) {\contour{white}{$1$}};
            \node[text=black] at (1.1,1.5) {\contour{white}{$1$}};
            
            \node[text=black] at (1.5,2.5) {};
            
        \end{tikzpicture}}}$  &
        \begin{tabular}{ |p{0.9cm}|p{1.2cm}|p{0.9cm}| } 
         \hline
         0.006 & 1.005 & 0.001 \\ \hline
         0.005 & 13.925 & 0.005 \\ \hline
         0.002 & 63.774 & 0.003 \\ \hline
         1.004 & 108.720 & 1.004 \\ \hline
         0.002 & 64.010 & 0.003 \\ \hline
         0.010 & 13.996 & 0.004 \\ \hline
         0.001 & 1.000 & 0.005 \\ \hline
        \end{tabular}
    \end{tabular}
    \caption{Some examples of the toric data predictions of our ResNet model, taken at various depths (i.e. number of Seiberg dualities performed). For each case, we show the corresponding $T^2$ tiling, the known toric data (i.e. the labels on which the model is trained), and the predictions from a particular run of the ResNet training. Multiplicities close to zero would correspond to empty points in the $\mathbb{Z}^2$ lattice. Upon rounding the predicted GLSM multiplicities to the nearest integer, all three examples display 100\% accuracy. The final example (at depth 6) was absent from the training or validation datasets. (We remind here the reader that we do not integrate out bivalent nodes, as to allow the ResNet to learn by itself that such operations preserve the superpotential algebra.)}
    \label{tab:resnet_examples}
\end{table}

\section{Conclusions and outlook}\label{ref:conclusions}

In this paper, we have begun exploring the applicability of machine learning methods in the classification of bipartite field theories on D3-branes probing toric CY$_3$ singularities and toric phases thereof. These theories admit a representation as dimer models, which, in turn, can be recast into Kasteleyn matrices; hence, this scenario naturally lends itself to a machine learning investigation. We have focussed on two simple examples, namely the countably infinite family of $\mathbb{Z}_m\times\mathbb{Z}_n$ orbifolds of the conifold, and the toric phases of the $Y^{6,0}$ singularity. For the former investigation, we first generated a dataset containing 1000 matrices for each $(m,n)$ we consider. Then, we constructed and trained a fully connected neural network for a regression analysis. Even after taking into account the simplicity of these brane tilings, the performance of both architectures is promising. In particular, we found that a fully connected architecture consisting of a 64-neuron dense layer with ReLu activation function and a batch size of 64 achieves R$^2=0.988$ in 450 epochs. We also saw that the network is particularly robust in excisions of various depth layers from within the training dataset, and slightly less so in perturbations in $(m,n)$ space. Although the achieved $R^2$ value is high, the corresponding MAE is not low enough to enable us to predict $(m,n)$ pairs confidently, for example by rounding the predictions to the nearest integer. This can, of course, partly be explained by our choice of simple network. We suspect that symmetry-informed neural networks would be more appropriate to tackle this problem.

For our second study, we instead employed a residual neural network to classify the toric phases of the $Y^{6,0}$ tilings. Again, we can observe a remarkable performance from the network. Tasked with classifying the toric phase identifier of a given Kasteleyn matrix, the ResNet achieved an accuracy of 100\% to machine precision. Finally, we trained the same ResNet as a regressor, as to predict the individual toric diagrams (including shape and GLSM multiplicities) of the phases of $Y^{6,0}$, and found a similarly positive performance. We were also able to observe that the inclusion of the \texttt{RandomFlip} layer, which enacts redefinitions of the paths $\gamma_{w,z}$ and vertex labelling against which the Kasteleyn matrices are defined, worsens the performance of both the classifier and regressor ResNets.

\begin{figure}[t]
    \centering
    \includegraphics[width=0.5\linewidth]{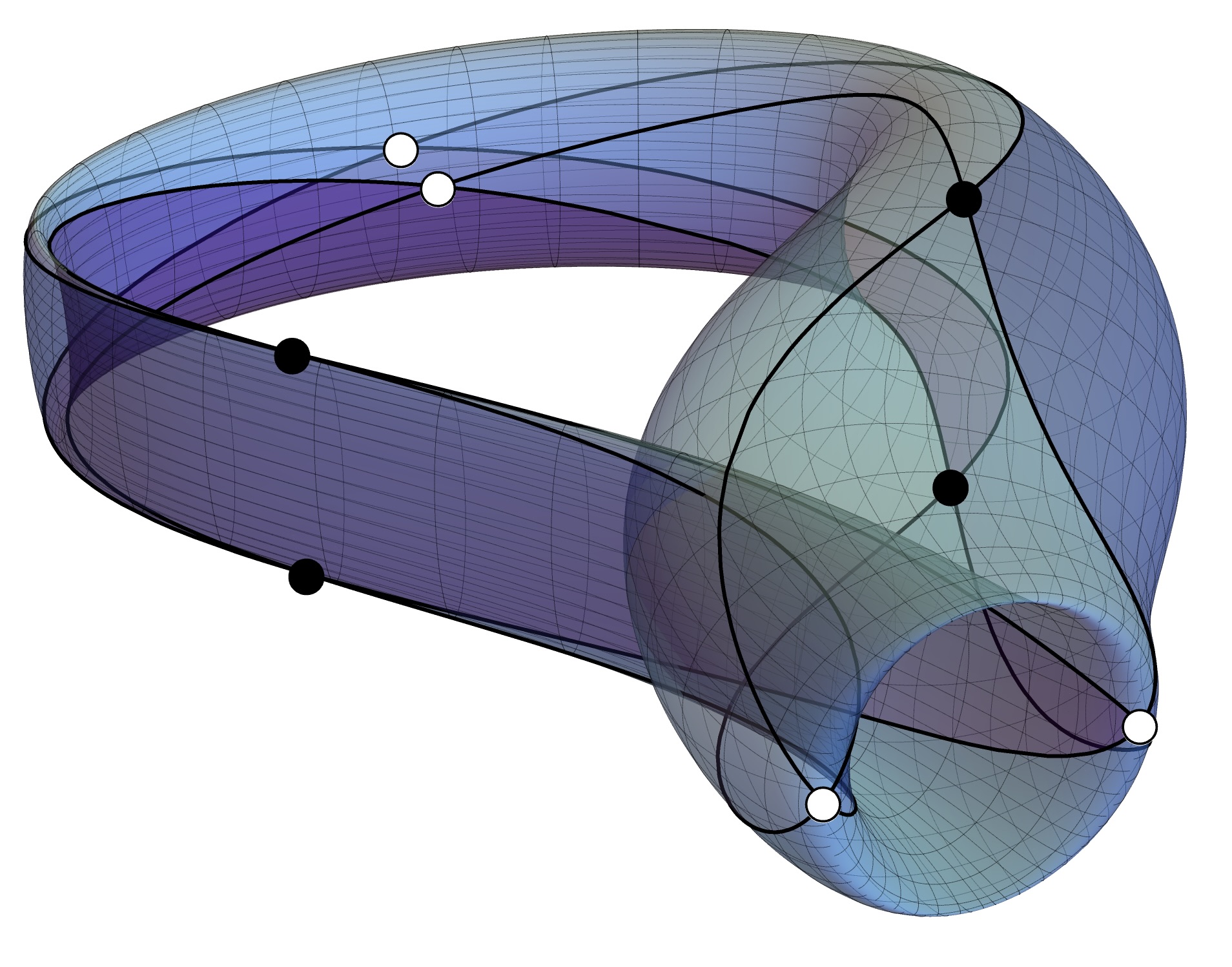}
    \caption{The tiling of the Klein bottle given by an orientifold projection of the $\mathbb{Z}_4\times\mathbb{Z}_2$ orbifold of the conifold.}
    \label{fig:bottle}
\end{figure}

These results strongly suggest that such machine learning architectures could also be successfully trained against certain straightforward generalisations and extensions, such as the inclusion of orientifolds of the conifold (shown in figure~\ref{fig:bottle}, and studied in \cite{Park:1999ep,Franco:2007ii,Garcia-Valdecasas:2021znu}), but perhaps also against much more generic datasets of dimer models, including for instance (pseudo) del Pezzo surfaces. Currently, and to the best of our knowledge, the main limitation impeding such an investigation actually concerns the data generation phase. Indeed, in order to satisfactorily train a supervised network, one is inevitably faced with the need to generate a large dataset of Kasteleyn matrices for geometries which may not admit an organisation as elegant as that of $\mathbb{Z}_m\times\mathbb{Z}_n$ orbifolds, and for which an iterative and systematic routine to generate arbitrarily large datasets may not be readily available.

An interesting augmentation of our analysis would be to employ reinforcement learning (RL) to not only deduce whether two tilings are Seiberg dual to each other, but also to explicitly identify the shortest sequence of Seiberg dualities connecting them. A similar model in the context of knot theory has recently been presented in \cite{Gukov:2020qaj}. The similarities between the two problems -- in both cases, one has an explicit representation (a brane tiling for BFTs, and a braid representation for knots) with a finite set of allowed moves (those of Seiberg duality for the tilings, and Reidemeister moves for knots, or equivalently Markov moves for braids) -- suggest that analogous methods could be successful for the classification of BFTs as well.

In this paper, we have considered 4d $\mathcal{N}=1$ theories arising on the worldvolumes of D3-branes probing singular toric CY$_3$ spaces. As amply discussed, these admit a diagrammatic description in terms of brane tilings. A natural extension of our investigation would then be to consider D1-branes and D$(-1)$-instantons probing singular toric CY 4- and 5-folds, which realise $\mathcal{N}=(0,2)$ and $\mathcal{N}=1$ gauge theories on their worldsheets and worldpoints, respectively. These can be described in terms of brane brick and hyper-brick models \cite{Franco:2015tya,Franco:2016tcm}, and enjoy interesting higher-order IR correspondences, such as trialities and quadrualities \cite{Gadde:2013lxa}, which a machine learning algorithm could be trained to recognise.

A perhaps more ambitious extension, and one we certainly hope to pursue soon, is the application of graph neural network (GNN) methods to the study of brane tilings. These would relieve the need for an explicit matrix representation and instead operate directly at the level of the bipartite graph on the torus (or on the infinite plane), and thus burke many of the technical issues we faced above, such as the need to generate large amounts of Kasteleyn matrices.

\section*{Acknowledgements}
The authors would like to thank David Vegh and Edward Hirst for useful discussions.
The work of PC is supported by a Mayflower studentship from the University of Southampton. The work of TSG is supported by the Science and Technology Facilities Council (STFC) Consolidated Grants ST/T000686/1 ``Amplitudes, Strings \& Duality'' and ST/X00063X/1 ``Amplitudes, Strings \& Duality''. The work of BS is supported in part by the STFC consolidated grant ST/T000775/1.

\appendix

\bibliographystyle{JHEP}
\bibliography{biblio}

\end{document}